\documentclass[prd,superscriptaddress,amsfonts,amssymb,amsmath,twocolumn,floatfix]{revtex4-2}

\usepackage{bm}
\usepackage{amsfonts}
\usepackage{latexsym}
\usepackage{graphicx}
\usepackage{amsmath}
\usepackage{palatino}
\usepackage{bigints}
\usepackage{mathpazo}
\usepackage{textcomp}
\usepackage{float}
\usepackage{booktabs}
\usepackage{dcolumn}
\usepackage{booktabs}
\usepackage{multirow}
\usepackage{hyperref}
\hypersetup{colorlinks,citecolor=blue}
\usepackage{amsmath}
\usepackage{xcolor}
\usepackage{orcidlink}
\usepackage{commath}
\usepackage{subcaption}
\def\nat{Nature }
\def\prl{Phys. Rev. Lett. }

\def\prd{Phys. Rev. D }

\def\apj{Astrophys. J. }
\def\apjl{Astrophys. J. Lett. }

\def\aap{Astron. Astrophys. }

\def\physrep{Phys. Rep. }

\newcommand{\bea}{\begin{eqnarray}}
\newcommand{\eea}{\end{eqnarray}}
\newcommand{\be}{\begin{equation}}
\newcommand{\ee}{\end{equation}}

 \catcode`\@=11 \@addtoreset{equation}{section}\catcode`\@=12

\newcommand{\R}{ {\mathbb R} }

\newcommand{\beq}[1]{\begin{equation}\label{#1}}
\newcommand{\eeq}{\end{equation}}

\newcommand{\bear}[1]{\begin{eqnarray}\label{#1}}
\newcommand{\bearr}[1]{\begin{eqnarray}\lal \label{#1}}
\newcommand{\ear}{\end{eqnarray}}

\newcommand{\p}{\partial}

\catcode`\@=11 \@addtoreset{equation}{section}\catcode`\@=12

\begin{document}

%\title{Circular geodesics in the field of dyon-like \\ dilatonic black hole}

\title{Stability Analysis of Circular Geodesics in Dyonic Dilatonic Black Hole Spacetimes}

\author{K.~Boshkayev\orcidlink{0000-0002-1385-270X}}
\email[Email: ]{kuantay@mail.ru}
%\affiliation{National Nanotechnology Laboratory of Open Type, Almaty 050040, Kazakhstan.}
\affiliation{Al-Farabi Kazakh National University, Al-Farabi av. 71, 050040 Almaty, Kazakhstan
}
\affiliation{Institute of Nuclear Physics, Ibragimova, 1, 050032 Almaty, Kazakhstan}
%\affiliation{Kazakh-British Technical University, Tole bi str., 59, Almaty, 050000, Kazakhstan}

\author{G.~Takey\orcidlink{0009-0006-6382-6776}}
\email[Email: ]{gulnaz-takey@mail.ru}
\affiliation{Al-Farabi Kazakh National University, Al-Farabi av. 71, 050040 Almaty, Kazakhstan
}

\author{V.D.~Ivashchuk\orcidlink{0000-0002-4153-2658}}
\email[Email: ]{ivashchuk@mail.ru}
\affiliation{Center for Gravitation and Fundamental Metrology, \\
Scientific Research Center of Applied Metrology Rostest, \\ 
Ozyornaya St. 46, Moscow 119361,  Russian Federation
}
\affiliation{Institute of Gravitation and Cosmology, Peoples' Friendship University of Russia (RUDN University),\\ 
Miklukho-Maklaya St. 6, Moscow 117198, Russian Federation}

\author{A.N.~Malybayev\orcidlink{0000-0003-2060-105X}}
\email[Email: ]{algis-malybayev@mail.ru}
\affiliation{Al-Farabi Kazakh National University, Al-Farabi av. 71, 050040 Almaty, Kazakhstan
}

\author{G.S.~Nurbakova\orcidlink{0000-0001-5999-8635}}
\email[Email: ]{guliya-nurbakova@mail.ru}
\affiliation{Al-Farabi Kazakh National University, Al-Farabi av. 71, 050040 Almaty, Kazakhstan
}

\author{A.~Urazalina\orcidlink{0000-0002-4633-9558}}
\email[Email: ]{y.a.a.707@mail.ru}
\affiliation{Al-Farabi Kazakh National University, Al-Farabi av. 71, 050040 Almaty, Kazakhstan
}
\affiliation{Institute of Nuclear Physics, Ibragimova, 1, 050032 Almaty, Kazakhstan}

\date{Received: date / Accepted: date}

%\maketitle

\begin{abstract}
This research investigates a non-extreme dyonic-like dilatonic charged black hole solution within a four-dimensional gravity model. This model incorporates two scalar (dilaton) fields and two Abelian vector fields, with interactions between the scalar and vector fields mediated by exponential terms involving two dilatonic coupling vectors. The solution is characterized by a dimensionless parameter $a$ (where $0 < a < 2$), which is specifically defined as a function of the dilatonic coupling vectors. The paper further explores solutions for timelike and null circular geodesics, which are crucial for understanding various astrophysical scenarios, including the quasinormal modes of different test fields in the eikonal approximation. For all values of  $a$ the innermost stable circular orbit (ISCO) are found by means of reducing the problem to the 
solution of fourth order polynomial equation.
\end{abstract}

\maketitle

\section{Introduction}

The detection of gravitational waves in 2016 \cite{2016PhRvL.116f1102A} sparked a surge of interest in the physics of stellar-mass black holes (BHs).  Einstein's century-old prediction of these ripples in spacetime led to an ambitious search, culminating in the construction of massive, exquisitely sensitive laser interferometers like LIGO, VIRGO, and others.  These instruments, paired with advanced detection and data analysis techniques \cite{2016PhRvL.116v1101A}, have opened a new window into the universe, allowing us to explore the most extreme phenomena in nature.

The heart of our Milky Way galaxy harbors a supermassive black hole (BH), a fact confirmed by meticulous observations and studies of stellar motion in this region \cite{1998ApJ...509..678G, 2000Natur.407..349G,2018A&A...615L..15G}. These surveys, revealing the dynamics of star clusters surrounding the galactic center and the BH's mass, provide invaluable data for testing the predictions of general relativity (GR). By carefully tracking the movement of stars in this extreme environment, scientists can probe the fabric of spacetime and refine our understanding of gravity's impact on the universe.

Capturing the elusive shadows of supermassive BHs at the heart of the M87 galaxy and our own Milky Way has been a monumental achievement, requiring an unprecedented level of scientific ingenuity and collaboration across the globe \cite{2019ApJ...875L...1E, 2022ApJ...930L..12E}. To achieve this, researchers effectively transformed our planet into a colossal radio telescope, coordinating observations from disparate groups of telescopes spanning all continents. The emergence of these BH shadows, along with their detailed analysis, provides compelling confirmation of the validity and power of GR. This remarkable achievement, coupled with decades of groundbreaking discoveries, underscores the paramount importance and enduring relevance of research in this field.

While GR reigns supreme in our current understanding of gravity, alternative theories abound, offering modified and extended descriptions of gravitational phenomena. These alternative frameworks often predict BHs with additional parameters, distinct from the familiar Schwarzschild, Reissner-Nordstr\"{o}m, and Kerr solutions \cite{Sotiriou:2006hs,2012PhR...513....1C,Astashenok:2014nua,Astashenok:2013vza,Astashenok:2017dpo,Capozziello:2019cav,Astashenok:2020qds}. However, current observational capabilities, with their inherent error margins, limit our ability to definitively distinguish these exotic BHs from their GR counterparts \cite{2022ApJ...930L..17E}. As our observational precision improves, we may be able to unravel the true nature of these celestial objects and definitively test the validity of different gravitational theories.

Observational data play a crucial role in constraining the parameters of BHs within various theoretical frameworks, including modified and extended theories of gravity \cite{2009PhRvD..79h4031P, 2022GReGr..54...44S, 2023Univ....9..147A,2023CQGra..40p5007V}. This wealth of observational information allows us to explore the properties of dyonic-like BHs with ``colored charges'' in the presence of scalar fields. This research delves into both the classic Schwarzschild and Reissner-Nordström BHs, as well as a variety of dilatonic BH models, each characterized by two distinct colored charges: both electric and magnetic ones. Through careful analysis of these diverse BH configurations, we aim to shed light on the intricate interplay between gravity, scalar fields, and the fundamental nature of these conjectural celestial objects. 

The dilaton, a hypothetical scalar field particle, interacts with the fabric of spacetime through coupling constants, which shape the metric and define a specific subclass of BHs. In certain limiting cases, these dilaton solutions smoothly transition to the well-known Schwarzschild and Reissner-Nordstr\"{o}m solutions, their behavior determined by the specific values of these coupling constants. While the dilaton is still elusive, its potential impact on the properties of BHs makes it an intriguing subject of research. Exploring the effects associated with its existence could reveal new insights into the complex interaction between gravity, scalar fields, and the nature of BHs themselves.

This paper considers the intriguing question of whether we can differentiate between ordinary astrophysical BHs and their dilatonic counterparts. Our primary goal is to analyze the motion of neutral test particles and photons in circular orbits around these celestial objects, focusing on BHs with variety of  ``color charges’’ (electric and magnetic ones) and coupling vectors. By scrutinizing the subtle differences in these orbital dynamics, we aim to reveal potential observational signatures that could distinguish dyonic-like dilatonic BHs from their more conventional counterparts. This investigation holds the potential to shed light on the mysterious properties of the dilaton and its impact.

To achieve our goal of distinguishing between ordinary and dilatonic BHs, we embark on a comprehensive analysis of orbital dynamics, tackling the following key problems:
- By applying the powerful framework of the Lagrange formalism, we derive the equations that govern the motion of test particles and photons in the spacetime warped by dilatonic BHs. 
- We rigorously calculate proper crucial quantities (the angular momentum, energy, effective potential of neutral test particles and photons) for both neutral massive test particles and photons, providing insights into the specific characteristics of orbits around dilatonic BHs. This includes determining the radii of the innermost stable circular orbits (ISCO), a key parameter that marks the boundary between stable and unstable orbits. 

The intriguing properties of dilatonic dyonic BHs, characterized by both electric and magnetic charges and  certain coupling vectors, have been extensively studied in recent years \cite{2015CQGra..32p5010A,ABI,2020JPhCS1690a2143B,MBI}. These investigations, utilizing a canonical scalar field, have yielded solutions by solving two master equations for the moduli functions. This analysis has revealed key physical parameters of these BHs, including their gravitational mass, scalar charge, Hawking temperature, area entropy, and parameterized post-Newtonian (PPN) parameters. While early research primarily focused on the intrinsic characteristics of these BHs \cite{2015CQGra..32p5010A,ABI,2020JPhCS1690a2143B}, a subsequent study of  Ref. ~\cite{MBI}  ventured into the realm of quasinormal modes, exploring the behavior of a massless test scalar field in the gravitational field surrounding a non-extremal dilatonic dyonic-like BH. This exploration opens new avenues for understanding the dynamics and potential observational signatures of these fascinating hypothetical celestial objects. 

The study of timelike and null geodesics, including circular orbits, is fundamental to  understanding a wide range of astrophysical phenomena.  
These trajectories play critical roles in the dynamics of accretion disks, the generation of quasiperiodic oscillations, the behavior of quasinormal modes of test fields in the eikonal approximation \cite{2019PhRvD..99l4042K}, and the formation of shadows cast by supermassive BHs \cite{2023PDU....4001178U,2023arXiv230400183L}.  By carefully analyzing these geodesics, researchers aim to discern potential differences between ordinary BHs and their dilatonic counterparts \cite{2009PhRvD..79h4031P}, ultimately contributing to our understanding of these enigmatic objects and the fundamental nature of gravity. 

This work breaks new ground by exploring the intricate paths of test particles in the  gravitational field of dilatonic BHs with a unique twist: the inclusion of two scalar fields and color electric and magnetic charges. This marks the first study of its kind, venturing into a realm previously unexplored.  While previous research has delved into the dynamics of geodesics within the contexts of standard BHs, naked singularities, and spinning deformed relativistic compact objects \cite{2021PhRvD.104h4009B,2016PhRvD..93b4024B,2013NCimC..36S..31B,2016IJMPA..3141006B}, this investigation focuses specifically on the unique characteristics of dyonic-like dilatonic BHs with their complex interplay of scalar fields and color charges.  Complementing this research, the geometric and thermodynamic properties of dilatonic BHs have been explored in the presence of linear and nonlinear electromagnetic fields \cite{2018PhRvD..98h4006P,2017EPJC...77..647H,2017PhLB..767..214H,2016EPJC...76..296H,2015PhRvD..92f4028H}. This collaborative effort promises to deepen our understanding of these fascinating objects and the fundamental nature of gravity.

A recent study in Ref.~\cite{CGP}, explored a  fascinating  realm of closed photon orbits within spherically symmetric static solutions of various gravity theories, including supergravity, Horndeski theory, and quintessence. These orbits, when unstable, trace a path known as a photon sphere (or an anti-photon sphere if stable). The study revealed a remarkable consistency: in all asymptotically flat solutions examined, those possessing a regular event horizon and an energy-momentum tensor satisfying the strong energy condition, only a single photon sphere exists outside the event horizon. This finding sheds light on the universal nature of photon sphere formation in these gravity theories, highlighting the fundamental role of the strong energy condition in shaping the geometry of spacetime around compact objects.

While recent studies have explored the intriguing features of dilatonic BHs 
~\cite{2022PhRvD.105l4056H,2020EPJC...80..654H,2010JHEP...03..100C,2008PThPS.172..161C}, others have focused on solutions to field equations that, in specific limits, describe ordinary BHs .~\cite{2022PhRvD.106h4041L,2022EPJC...82..771S,2022EPJP..137..222T}. These investigations, including those examining geodesics in the vicinity of astrophysical BHs ~\cite{2006ChPhL..23.1648Z,2015SerAJ.190...41B, 2018arXiv180500295S,2023Symm...15..329B,2022PhRvD.105l4009H}, provide valuable insights into the rich landscape of gravitational phenomena.  Additionally, a number of works have explored the motion of test particles and photons around static \cite{2006ChPhL..23.1648Z,2015SerAJ.190...41B, 2018arXiv180500295S,2023Symm...15..329B,2022PhRvD.105l4009H}  and rotating \cite{2016PhRvD..94b4010S,2015PhRvD..92j4027F}  dilatonic BHs within various theories of gravity. However, to the best of our knowledge, no previous studies, but that of Refs. ~\cite{IMNT,BSIU}, have considered  the dynamics of geodesics around dilatonic BHs featuring both two scalar fields and two vector fields. This unexplored territory represents a significant frontier in understanding the complex interplay between these fields and their impact on the geometry of spacetime.

In the realm of classical physics, gravity's influence is inextricably linked to mass, leaving charge unaffected.  This separation of forces leads to a distinct lack of interplay between gravitational and electromagnetic fields.  However, within the framework of GR, the picture becomes richer and more intricate. Gravity, in addition to its dependence on mass, can be generated by a rotating object's kinetic energy (the Lense-Thirring effect), the presence of electric or magnetic charges, the energy associated with their electromagnetic fields, and even other types of fields. This interconnectedness implies that gravity exerts its influence on the motion of both neutral and charged particles in GR.  These effects become particularly pronounced in the intense gravitational fields surrounding BHs and neutron stars, where the curvature of spacetime becomes significant.
       
Interestingly, neutral and charged particles exhibit distinct behavior under the influence of gravity. While neutral particles follow geodesic paths in the absence of external forces, charged particles can deviate significantly from these geodesic trajectories. The curvature of spacetime itself, generated by massive and compact objects, can dramatically alter the paths of charged particles. Moreover, the presence of strong magnetic fields introduces the Lorentz force, further deviating charged particles from geodesic motion. While magnetic fields can influence the motion of neutral test particles, this influence is typically less pronounced. Similarly, strong electric fields exert Coulomb forces on charged particles, leading to deviations from geodesic motion. This effect becomes particularly relevant in astrophysics, where charged particles navigate the intense magnetic and electric fields surrounding neutron stars or within astrophysical jets.

A comprehensive overview of the interplay between rotation, the cosmological constant, and magnetic fields on the motion of charged and neutral particles within accretion disks surrounding Kerr BH candidates is presented in  Ref.~\cite{2020Univ....6...26S}. This work explores realistic astrophysical scenarios around compact objects, encompassing the characteristics of thin and thick, neutral and charged accretion disks, quasiperiodic oscillations, and relativistic jets.

This paper examines a specific solution for a dyonic-like dilatonic BH, previously derived in \cite{MBI}, and deals with particular solutions for null and timelike geodesics within this spacetime. The BH solution under consideration arises from a gravitational model featuring two ``neutral'' (real-valued) scalar fields and two Abelian gauge fields.  

The study of spherically symmetric solutions in gravitational models is a topic of ongoing interest, as exemplified by numerous investigations, including those found in Refs.~\cite{BS,1988NuPhB.298..741G,1991PhRvD..43.3140G,1992PhRvD..45.3888G} and others.  These solutions often arise in models incorporating scalar fields and antisymmetric forms. In our analysis, we adopt a methodology similar to that employed in Ref. ~\cite{2011PhRvD..83b4021P}, where the motion of neutral test particles in the gravitational field of a Reissner-Nordstr\"{o}m BH was explored. This approach provides a solid foundation for investigating the dynamics of particles in more complex and intriguing spacetimes.

This paper is structured as follows.
In Section \ref{sec:metrics} we introduce charged BH solutions characterized by two scalar (dilaton) fields and two Abelian vector fields. The key features of these solutions are discussed in detail.
Section \ref{sec:geod_motion} focuses on the analysis of geodesics followed by neutral test particles and photons. We derive expressions for energy, angular momentum, and effective potential. We also scrutinize their behavior and analyze the stability of circular geodesics. In Section \ref{sec:examples} certain examples of solutions are explored. Finally, in Section \ref{sec:conclusion}, we summarize our results and discuss potential lines for future research.

%%%%%%%%%%%%%%%%%%%%%%%%%%%%%%%%%%%%%%%%%%%%%%%%%%%%%%%%%%%%%%%%
\section{Dyonic-like black hole solution} \label{sec:metrics}
%%%%%%%%%%%%%%%%%%%%%%%%%%%%%%%%%%%%%%%%%%%%%%%%%%%%%%%%%%%%%%%%

The action of a model incorporating two scalar fields, two 2-forms, and dilatonic coupling vectors is described by

\bear{i.1}
 S= \frac{1}{16 \pi G}  \int d^4 x \sqrt{|g|}\biggl\{ R[g] -
  g^{\mu \nu} \p_{\mu} \vec{\varphi}  \p_{\nu} \vec{\varphi}
 \qquad \qquad   \nonumber \\
 - \frac{1}{2} e^{2 \vec{\lambda}_1 \vec{\varphi}} F^{(1)}_{\mu \nu} F^{(1)\mu \nu }
 - \frac{1}{2} e^{2 \vec{\lambda}_2 \vec{\varphi}} F^{(2)}_{\mu \nu} F^{(2) \mu \nu}
 \biggr\},\quad 
\ear
where $g= g_{\mu \nu}(x)dx^{\mu} \otimes dx^{\nu}$ is the metric,  $|g| =   |\det (g_{\mu \nu})|$, $\vec{\varphi} =  (\varphi^1,\varphi^2)$ is the vector (set) of two scalar fields belonging to ${\R}^2$,  $F^{(i)} = dA^{(i)}  =  \frac{1}{2} F^{(i)}_{\mu \nu} dx^{\mu} \wedge dx^{\nu}$ is the $2$-form with $A^{(i)} = A^{(i)}_{\mu} dx^{\mu}$, $i =1,2$; $G$ is the gravitational constant, $\vec{\lambda}_1 = (\lambda_{1i}) \neq \vec{0}$,  $\vec{\lambda}_2 = (\lambda_{2i}) \neq \vec{0}$ are the dilatonic coupling  vectors  which obey 
 
\beq{i.1a}
\vec{\lambda_1} \neq  - \vec{\lambda_2}, 
\eeq
and $R[g]$ is the Ricci scalar. Here, and in what follows, we set $G =c =1$ (where $G$ is the gravitational constant and $c$ is the speed of light in vacuum.)

We consider a  dyonic-like BH solution \cite{MBI} to the  field equations corresponding to the action (\ref{i.1}) which is defined on the (oriented) manifold 
\beq{i.2}
 {\cal M }  =   \R \times(2\mu, + \infty)  \times S^2   ,
\eeq
and has the following form 
\bear{i4.9}
 ds^2 &=& H^{a}\biggl\{-H^{-2a} \left( 1 - \frac{2\mu}{R} \right) dt^2 \nonumber\\
  &\color{white}+& \quad\qquad\qquad +\frac{dR^2}{1 - \frac{2\mu}{R}} + R^2 d \Omega^2 \biggr\},
 \\  \label{i4.10}
 \varphi^i &=& \nu^i \ln H , 
\ear
with the 2-form defined by
%\vspace{-0.5cm}

\beq{i4.10em}
 F^{(1)} = \frac{Q_1}{H^2 R^2} dt \wedge dR, \quad
 F^{(2)}  = Q_2 \tau,    
\eeq
where $Q_1$ is the (color) electric charge and $Q_2$ are the (color) magnetic charge.  

Here the extremality parameter is denoted by $\mu > 0$,  the metric on the unit sphere $S^2$ is given by $d \Omega^2 = d \theta^2 + \sin^2 \theta d \phi^2$, where $0 < \theta < \pi$ and $0 < \phi < 2 \pi$, and the standard volume form on $S^2$ is $\tau = \sin \theta d \theta \wedge d \phi$. The moduli function is defined as

\beq{i4.7}
     H = 1 + \frac{P}{R}.
\eeq
   The parameter $P > 0$ obeys
\beq{i4.8a}
     P (P + 2 \mu) = \frac{1}{2} Q^2 ,
\eeq
   or, equivalently
\beq{i4.8b}
     P= -\mu + \sqrt{\mu^2 + \frac{1}{2} Q^2}.
\eeq

All the remaining parameters of the solution are defined as follows 

\bear{i4.10a}
   a &=&  \frac{  ( \vec{\lambda}_1 + \vec{\lambda}_2 )^2}{ \Delta },
 \\  \label{i4.10n}
     \Delta &\equiv& \frac{1}{2}  (\vec{\lambda}_1 + \vec{\lambda}_2)^2 +
      \vec{\lambda}_1^2 \vec{\lambda}_2^2 - (\vec{\lambda}_1 \vec{\lambda}_2)^2,
\\ \label{i2.BB} 
\nu^i &=& \frac{ \lambda_{1i} \vec{\lambda}_2 ( \vec{\lambda}_1 + \vec{\lambda}_2 )
    - \lambda_{2i} \vec{\lambda}_1 ( \vec{\lambda}_1 + \vec{\lambda}_2 )}{ \Delta }, \qquad
  \ear
$i = 1,2$ and
 \beq{i4.8c}
    Q_1^2  = \frac{ \vec{\lambda}_{2}( \vec{\lambda}_1 + \vec{\lambda}_2 )}{2 \Delta} Q^2, \quad
    Q_2^2  = \frac{ \vec{\lambda}_{1}( \vec{\lambda}_1 + \vec{\lambda}_2 )}{2 \Delta} Q^2. 
 \eeq

Here, the following additional constraints on dilatonic coupling  vectors are imposed      
     \beq{i4.8bdd}
        \vec{\lambda}_{1}( \vec{\lambda}_1 + \vec{\lambda}_2 ) > 0,  \qquad 
         \vec{\lambda}_{2}( \vec{\lambda}_1 + \vec{\lambda}_2 ) > 0. 
     \eeq
Note that the restrictions  (\ref{i4.8bdd}) imply inequalities 
 $\vec{\lambda}_s \neq \vec{0}$, $s = 1,2$, and  (\ref{i.1a}).
 
It can be readily verified that   
  \beq{i.18BD}
       \Delta > 0,
  \eeq
is valid for   $\vec{\lambda}_1 \neq -  \vec{\lambda}_2$.
Indeed,  in this case we have the sum of two non-negative terms in (\ref{i4.10n}):  $\frac{1}{2} (\vec{\lambda}_1 + \vec{\lambda}_2)^2 > 0$ and 
    
\beq{i.18BC}
     C = \vec{\lambda}_1^2 \vec{\lambda}_2^2 - (\vec{\lambda}_1 \vec{\lambda}_2)^2 \geq 0,
\eeq
  
because of the Cauchy--Schwarz inequality. Here, $C = 0$ if and only if vectors $\vec{\lambda}_1$ and $\vec{\lambda}_2$ are collinear. Relation (\ref{i.18BC}) implies 
   \beq{i.18a}
     0 < a \leq 2.
   \eeq
For non-collinear vectors  $\vec{\lambda}_1$ and $\vec{\lambda}_2$ we find  $0 < a < 2$ , and for collinear ones we get $a = 2$ .

We note that due to relations (\ref{i4.8bdd}) and (\ref{i.18BD}), the quantities of charges $Q_s$ are well defined (up to signs). 
   
The gravitational mass is given by a relation where parameters  $\mu$, $a$ and $P$ are used  ~\cite{MBI},:
\beq{i.18bb}
     M = \mu + \frac{a}{2} P.
\eeq
The mass $M$ and charge $Q$ obeys 
the inequality ~\cite{MBI}
\beq{i.18QM}
     \frac{Q^2}{M^2} < \frac{8}{a^2} .
\eeq

Eqs.~\eqref{i4.10a}-\eqref{i4.8c} imply the following relations  
\beq{i5.1simQ2}
   \vec{\nu}^2 = \frac{(\vec{\lambda}_1 + \vec{\lambda}_2)^2
    (\vec{\lambda}_1^2 \vec{\lambda}_2^2 - (\vec{\lambda}_1 \vec{\lambda}_2)^2 )}{ \Delta^2}
    = \frac{a (2-a)}{2},\qquad
\eeq
\beq{i5.4Q}
    Q_1^2 +  Q_2^2 = \frac{a}{2} Q^2.
\eeq

It follows from ~\eqref{i5.1simQ2} and ~\eqref{i5.4Q} that for vanishing $a\rightarrow 0$ we get $\nu^i\rightarrow 0$, $Q_i\rightarrow0$ and $\varphi^i\rightarrow 0$ and the line element Eq.~\eqref{i4.9} reduces to the Schwarzschild metric. 

Calculating the scalar curvature for the metric $ds^2 = g_{\mu \nu} dx^{\mu} dx^{\nu}$  from (\ref{i4.9}) yields the following result  \cite{MBI}:
\beq{i4.8R}
          R[g]  =  \frac{a(2-a)P^2 (R - 2 \mu)}{2 R^{4 - a} (R+ P)^{1 + a}}.
\eeq   
For the Schwarzschild $(a=0)$ and Reissner--Nordstr\"{o}m $(a=2)$ solutions, one immediately obtains $R[g]=0$. 

Table~\ref{tab:amuP} presents $\mu$ and $P$ in terms of $M$ and $Q$ for certain values of $a$.

\begin{table}
\centering
\setlength{\tabcolsep}{1.em}
\renewcommand{\arraystretch}{1.1}
\begin{tabular}{lcc}
\hline
\hline
   &                                           &       \\
$a$    & $\mu$ & $P$  \\
    &                                           &       \\
\hline
   &                                           &       \\
0.00 & $M$                                       & --   \\
    &                                           &       \\
0.25 &  $\frac{7M}{6} - \frac{\sqrt{8M^2 + 3Q^2}}{12\sqrt{2}}$ &  $-\frac{4M}{3} + \frac{\sqrt{2}}{3}\sqrt{8M^2 + 3Q^2}$     \\
    &                                           &       \\
0.50 & $\frac{3M}{2}-\frac{1}{4}\sqrt{4M^2+Q^2}$ & $-2M+\sqrt{4M^2+Q^2}$\\
    &                                           &       \\
0.75 & $\frac{5M}{2} - \frac{3 \sqrt{8M^2 + Q^2}}{4\sqrt{2}}$ &  $-4M + \sqrt{2} \sqrt{8M^2 + Q^2}$     \\
      &                                           &       \\
1.00 & $M-\frac{Q^2}{8M}$                        & $\frac{Q^2}{4M}$ \\
    &                                           &       \\
1.25 & $-\frac{3M}{2} + \frac{5 \sqrt{8M^2 - Q^2}}{4\sqrt{2}}$ & $4M - \sqrt{2} \sqrt{8M^2 - Q^2}$       \\
      &                                           &       \\
1.50 & $-\frac{M}{2}+\frac{3}{4}\sqrt{4M^2-Q^2}$ & $2M-\sqrt{4M^2-Q^2}$\\
    &                                           &       \\
1.75 & $ -\frac{M}{6} + \frac{7 \sqrt{8M^2 - 3Q^2}}{12\sqrt{2}}$  & $\frac{4M}{3} - \frac{\sqrt{2} }{3}\sqrt{8M^2 - 3Q^2}$      \\
    &                                           &       \\
2.0 & $\sqrt{M^2-\frac{Q^2}{2}}$                & $M-\sqrt{M^2-\frac{Q^2}{2}}$  \\
   &                                           &       \\
\hline
\end{tabular}
\caption{Values of parameters $\mu$ and $P$ in terms of $M$ and $Q$ depending on $a$. Note that for vanishing $Q$ one obtains $P=0$ and $\mu=M$. At the same time, for vanishing $a$, one recovers the Schwarzschild metric, and $\mu=M$ and $P$ disappears in the metric}
\label{tab:amuP}
\end{table}

%%%%%%%%%%%%%%%%%%%%%%%%%%%%%%%%%%%%%%%%%%%%%%%%%%%%%%%%%%%%%%%%%%%%%%%%%%%%%%%%%%%%%%%%%%%%%%%%%%%%%%%%%%%%%%%%%%%%

\section{Geodesic solutions} \label{sec:geod_motion}
%%%%%%%%%%%%%%%%%%%%%%%%%%%%%%%%%%%%%%%%%%%%%%%%%%%%%%%%%%%%%%%%%%%%%%%%%%%%%%%%%%%%%%%%%%%%%%%%%%%%%%%%%%%%%%%%%%%%

The study of geodesics is fundamental to understanding the motion of test particles within the gravitational field of dilatonic BHs. Such analysis offers valuable insight into the nature of these objects and their influence on the surrounding spacetime. 

These geodesics are derived from the Lagrangian:
\be \label{eq:Lag}
    \mathcal{L} = \frac{1}{2}  g_{\alpha \beta}(x) \dot{x}^{\alpha}\dot{x}^{\beta}.
\ee

The geodesic equations are equivalent to the Euler-Lagrange equations:
\be \label{eq:ELeqs}
    \frac{d}{d\tau}\left(\frac{\partial \mathcal{L}}{\partial \dot{x}^{\alpha}}\right) - \frac{\partial \mathcal{L}}{\partial x^{\alpha}} = 0.
\ee

Here $\dot{x}^{\alpha}=dx^{\alpha}/d\tau=u^{\alpha}$ is the 4-velocity vector, e.g. of a test particle, moving along  the curve $x^{\alpha}(\tau)$. 
The variable $\tau$ is affine parameter in the case of null  geodesics and the proper time for a massive point-like particle which moves along  timelike geodesics, respectively, $\alpha=0, 1, 2, 3$. 

The generalized momentum, $p_{\alpha} = g_{\alpha \beta}(x) \dot{x}^{\beta}$, derived from the Lagrangian ~\eqref{eq:Lag}, is normalized as follows:
\be \label{eq:normalization}
     g^{\alpha \beta}(x) p_{\alpha} p_{\beta} =
    g_{\alpha \beta}(x)  u^{\alpha} u^{\beta} = -k.
\ee

Here, $k = -1, 0, 1$ denotes spacelike, null, and timelike geodesics, respectively. For a general value of $k$, the quantity $(-k/2)$ represents the conserved energy for the Lagrangian ~\eqref{eq:Lag}. For brevity, we will not consider spacelike geodesics in this paper.

Without loss of generality we consider null and timelike geodesics confined to the equatorial plane ($\theta = \pi/2$).  For the metric given by Eq. ~\eqref{i4.9}, the Lagrangian is:
\be \label{eq:Lagrangian}
     \mathcal{L}
    = \frac{1}{2} H^a\left[-H^{-2a}\left(1-\frac{2\mu}{R}\right)\dot{t}^2 + \frac{ \dot{R}^2}{1-\frac{2\mu}{R}} + R^2 \dot{\phi}^2\right].
\ee

Given cyclic coordinates $t$ (time) and $\phi$ (angle), the system possesses conserved quantities described by the integrals of motion
\be \label{eq:E_tilda}
    \tilde{E} = H^{-a}\left(1-\frac{2\mu}{R}\right)\dot{t},\quad
    \tilde{L} = H^a R^2 \dot{\phi},
\ee

related for $k=1$ to the total energy $E=\tilde{E} m$ and angular momentum $L=\tilde{L}m$, respectively, of a test (neutral) point-like particle of mass $m$. 

Now we deal with Eq.~\eqref{eq:normalization}. For the  line element from Eq.~\eqref{i4.9} it reads as follows 
\be
    -H^{-a}\left(1-\frac{2\mu}{R}\right)\dot{t}^2 + 
    \frac{ H^a \dot{R}^2}{1-\frac{2\mu}{R}} + H^a R^2 \dot{\phi}^2 = -k. 
\ee
Using Eqs. \eqref{eq:E_tilda}, this relation simplifies to the following differential equation:
\be  \label{eq:EqWithEL}
    -\frac{H^a E^2}{m^2 \left(1-\frac{2\mu}{R}\right)} + \frac{H^a \dot{R}^2}{1-\frac{2\mu}{R}} + \frac{H^{-a} L^2}{m^2 R^2} = -k,
\ee
which can be presented in a compact form:
\be \label{eq:EqForR}
    \dot{R}^2 + V^2 = \frac{E^2}{m^2},
\ee
 by using the effective potential
\be \label{eq:effective_potential}
    V = \sqrt{H^{-2a}\left(1-\frac{2\mu}{R}\right)\left(H^a k + \frac{L^2}{m^2 R^2}\right)}.
\ee

The Lagrange equation  for the radial coordinate $R$ 
may be presented in the following form
 
 \be \label{eq:EqForRtrue}
   \ddot{R} + V \frac{\partial V}{\partial R} = 0.
\ee

\begin{figure}%[ht]
\includegraphics[width=\columnwidth]{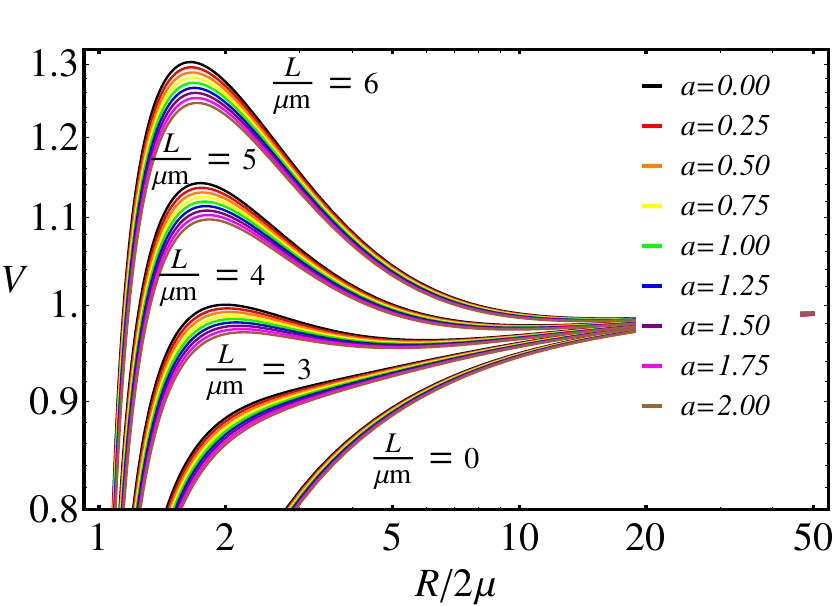}
\caption{The effective potential of test particles for $L/(\mu m)=0, 3, 4, 5, 6$ with different $a=0, 0.25, 0.5, 0.75, 1, 1.25, 1.5, 1.75, 2$ as a function of the dimensionless/normalized radial coordinate $R/(2\mu)$ \cite{BSIU}. As one may notice that for decreasing $L$, the minimum of the effective potential shifts from right to left. }
\label{fig:Veff_pl}
\end{figure}

For  $\dot{R} \neq 0$, it  follows just from Eq.~\eqref{eq:EqForR}. 

For circular trajectory with $\dot{R} = 0$, the radial equation ~\eqref{eq:EqForRtrue} reads
\be \label{eq:EqForR0}
    \frac{\partial V}{\partial R }= 0,
\ee
while equation (\ref{eq:EqForR}) takes the following form
\be \label{eq:VEm}
     V^2 = \frac{E^2}{m^2}.
\ee

Eq. ~\eqref{eq:EqForR0}  is not a direct consequence of Eq. ~\eqref{eq:EqForR} and requires separate consideration.

We now focus on timelike geodesics $(k=1)$. In Figure \ref{fig:Veff_pl}, we illustrate the behavior of the effective potential as a function of $R/\mu$ for a fixed value of $Q/\mu = 0.6$ (which is roughly $P/\mu = 0.0863$) and different values of $L/(\mu m)$. The cases of $a = 0$, $a = 1$, and $a = 2$ formally correspond to the Schwarzschild, Sen \cite{1992PhRvL..69.1006S}, and Reissner--Nordstr\"{o}m solutions, respectively. It is clearly seen that the maximum of the effective potential for $a = 0$ exceeds the maximums for other cases of $a$.

\begin{figure}%[ht]
\includegraphics[width=\columnwidth]{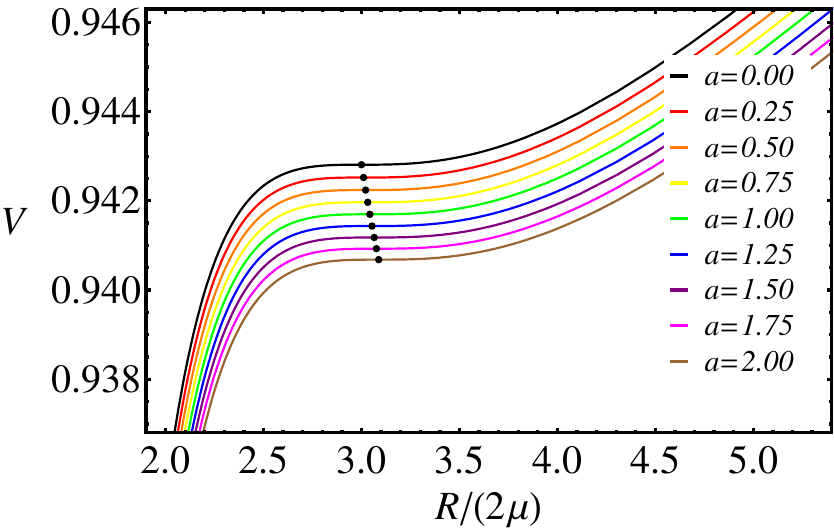}
\caption{The effective potential of test particles with  $L_{ISCO}$ for different $a$=0, 0.25, 0.5, 0.75, 1, 1.25, 1.5, 1.75, 2 as a function of the dimensionless (radial) coordinate $R/(2\mu)$ \cite{BSIU}. The dots represent the inflection points ($R_{ISCO}/(2\mu)$, $V_{ISCO} = V(R_{ISCO}/(2\mu))$)}
\label{fig:Visco}
\end{figure}

In Fig.~\ref{fig:Visco}, the effective potential at fixed $L =L_{ISCO}$ (see below) is illustrated as a function of the dimensionless radial coordinate $R/(2\mu)$ for different $a$=0, 0.25, 0.5, 0.75, 1, 1.25, 1.5, 1.75, 2. 

Here dots show the inflection points. As shown, the locations of these points, expressed in terms of $R$ and $\mu$, differ from those of standard Schwarzschild and Reissner--Nordstr\"{o}m BHs.

\subsection{Circular geodesics} \label{ssec:circ_orb}

Here, we consider circular motions, which are described by condition: $\dot{R} = 0$, so $V = E/m$. 

The first derivative of the effective potential with respect to $R$ reads:
\bea 
\label{eq:derivative_of_EP}
    \frac{\partial V}{\partial R} &=& \left[2R^3 H^{2+a} V\right]^{-1}\Big[H^a (a P(R-2\mu)\nonumber\\
    &+&2\mu(P+R))+\frac{2L^2}{m^2 R^2}\Big\{R(3\mu-R)\nonumber\\
    &+&P(R(a-1)+\mu(3-2a))\Big\}\Big].
\eea

The radial equation ~\eqref{eq:EqForR0} implies

\be \label{eq:L_of_R}
    \frac{L^2}{m^2} = \frac{H^a R^2 \left[aP(R-2\mu) + 2\mu(P+R)\right]}{2 \Delta_0},
\ee
where
\be \label{Delta}
 \Delta_0 =  R^2 + ((1 - a)P - 3 \mu)R + (2a-3) P \mu,
\ee
After substitution of Eq. ~\eqref{eq:L_of_R}  to Eq. ~\eqref{eq:effective_potential} 
and Eq. ~\eqref{eq:VEm} we get

\be \label{eq:E_of_R}
    \frac{E^2}{m^2} = \frac{H^{-a} (R-2\mu)^2 \left[2R-P(a-2)\right]}{2 \Delta_0}.
\ee

\begin{figure}[ht]
\includegraphics[width=\columnwidth]{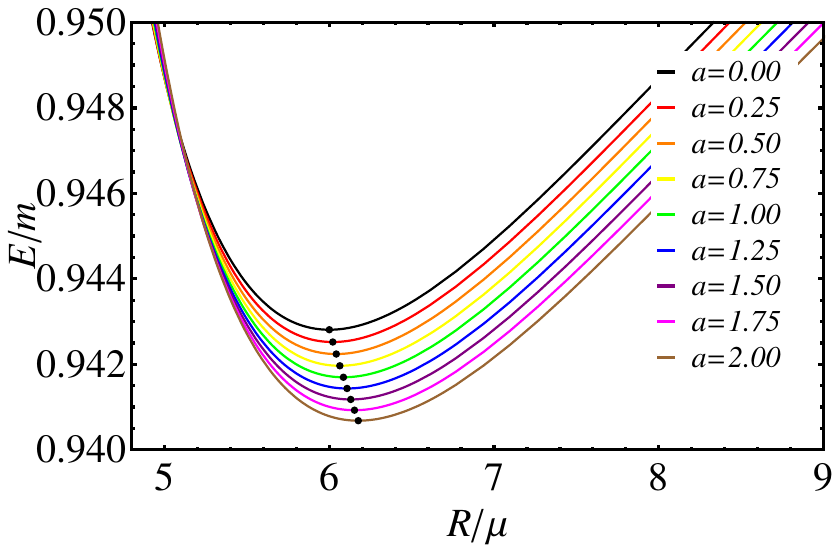}
\caption{Energy $E/m$ of test particles as a function of $R/\mu$ for cases $a$=0, 0.25, 0.5, 0.75, 1, 1.25, 1.5, 1.75, 2, if $Q/\mu=0.6$ \cite{BSIU}. The dots represent the values of $R_{ISCO}/\mu$ and the corresponding $E_{ISCO}/\mu$.}
\label{fig:E_of_R(a=0-2)}
\end{figure}

\begin{figure}[ht]
\includegraphics[width=\columnwidth]{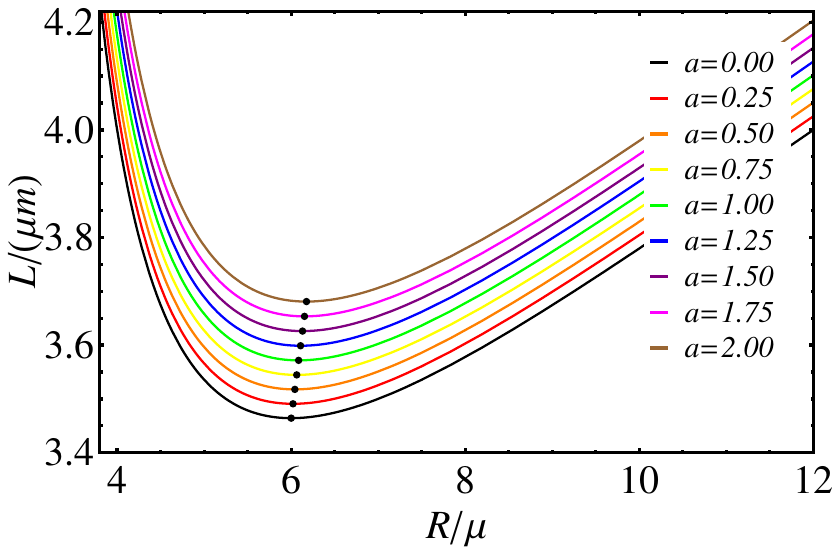}
\caption{Orbital angular momentum $L/(\mu m)$ of test particles as a function of $R/\mu$ for cases $a$=0, 0.25, 0.5, 0.75, 1, 1.25, 1.5, 1.75, 2, if $Q/\mu=0.6$ \cite{BSIU}. The dots represent the values of $R_{ISCO}/\mu$ and the corresponding $L_{ISCO}/\mu$.}
\label{fig:L_of_R(a=0-2)}
\end{figure}

Figures ~\ref{fig:E_of_R(a=0-2)} and \ref{fig:L_of_R(a=0-2)} plot  $E/m$ and $L/(\mu m)$ as functions of $R/\mu$ for a fixed value of $Q/\mu = 0.6$. 

The distinctions between the curves for different values of $a$ are larger for larger values of $Q/\mu$.

Equations~\eqref{eq:L_of_R} and \eqref{eq:E_of_R} reveal that for timelike geodesics, motion occurs only under the following condition: $\Delta_0 >0$, or, if  
\bea 
\label{eq:r_0}
  R>R_{0} &\equiv& \frac{1}{2} \Big[P(a-1)+3\mu \nonumber\\
   &+& \sqrt{(P(1-a)-3\mu)^2 - 4P\mu(2a-3)}\Big],\qquad\quad
\eea
where the radius of the photonic sphere is effectively given by the value of $R_0$, obeying $R_0 > 2 \mu$.

\begin{figure}[ht]
\includegraphics[width=\columnwidth]{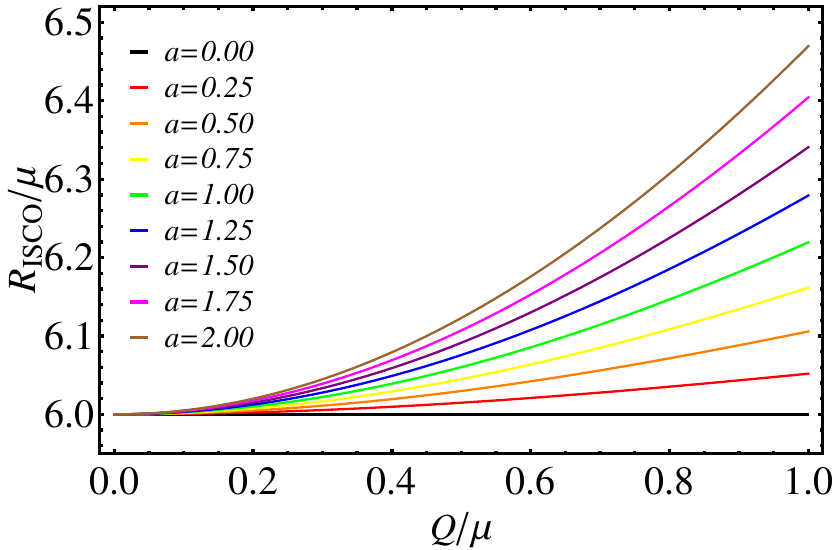}
\caption{The normalized ISCO radii $R_{ISCO}/\mu$ as a function of $Q/\mu$   \cite{BSIU}.}
\label{fig:xiscoq}
\end{figure}

It follows from Eq.~\eqref{eq:L_of_R} and Eq. ~\eqref{eq:E_of_R} that 
\be \label{eq:dL_of_dR}
     \frac{1}{m^2} \frac{dL^2}{dR} = \frac{ R^2 (1+P/R)^a}{\Delta_1^2} W, 
\ee
\be \label{eq:dE_of_dR}
     \frac{1}{m^2}\frac{dE^2}{dR} = \frac{(1-2\mu/R)(1+P/R)^{-a}}{\Delta_1^2} W,
\ee
with
\be \label{eq:Delta_1}
     \Delta_1 =  \frac{2}{R (R + P) (R - 2 \mu)} \Delta_0,
\ee
and 
\be \label{eq:W}
     W =  \frac{(2 \mu)^5}{R^3 (R + P)^3 (R- 2 \mu)^2} F,
\ee
where 

\begin{eqnarray}
       \small
       F  = (2 a p +2)x^4+ (6a(1-a)p^2+(6-12a)p-6)x^3   \nonumber \\
            + p(2(a-2)(a-1)a p^2 + 3(5a^2-9a+2)p +12a-18)x^2  \nonumber \\
        - p^2 ((a-2)(4a^2 - 7a +1)p +9a^2-25a+18)x \qquad \nonumber \\
         + (a-2)(a-1)(2a-3)p^3,  \qquad  \qquad 
         \label{eq:F} 
\end{eqnarray}
where $x = R/(2\mu)$, $p = P/(2 \mu)$.

\subsection{Innermost stable circular orbits} \label{ssec:isco}

The innermost stable circular orbit (ISCO) is a crucial concept in the study of objects around BHs. It is particularly significant in the context of accretion disks, as it determines the inner edge of these spinning structures. Any object that enters the ISCO is inevitably pulled into the BH. To maintain their integrity, accretion disks must have a radius greater than the ISCO. The location of the ISCO, along with characteristics such as the size, temperature, and luminosity of the accretion disk, provides valuable insights into the behavior of BHs and their surrounding environment.

It may be verified that the following is valid
\begin{equation}
\frac{\partial^2 V^2}{\partial R^2} = 2V \frac{\partial^2 V}{\partial R^2} = \frac{(1-2\mu/R)(1+P/R)^{-a}}{\Delta_1} W.
 \label{d2VdR2}
\end{equation}
Here $\Delta_1 > 0$, since $\Delta > 0$, see Eq.~\eqref{eq:Delta_1}.

By using the relation 
\begin{equation}
\frac{\partial^2 V}{\partial R^2} = 0, 
 \label{d2VdR20}
\end{equation}
or, equivalently, by setting the derivatives of Eqs.~\eqref{eq:dL_of_dR} and \eqref{eq:dE_of_dR} to zero, one can find a master equation for $x_{isco} = R_{isco}/(2 \mu)$.

\begin{eqnarray} 
       F(x_{isco}) = 0,   
  \label{master}     
\end{eqnarray}
which 

{\bf Proposition 1.} 
{\it Let us consider fourth order polynomial equation (\ref{master}),
where real parameters $a$ and $p$  obey inequalities: 
\begin{equation} 
      0 < a < 2,  \quad p > 0.       
 \label{ap}               
\end{equation}     
Then,  the  equation (\ref{master}) has one and only one solution  $x  = x_{*} =  x_{*}(a,p)$ which satisfies the relation 
\begin{equation} 
      x_{*} > 1;  \qquad             
     \label{xb1}     
\end{equation}              
moreover,  this solution obeys the following bound
    \begin{equation}
    x_{*}  >  x_0 \equiv \frac{1}{2} [(a - 1)p  + 3/2 + \sqrt{d}],  
     \label{xbx0}    
   \end{equation}                  
where $d = (1 - a)^2 p^2 + (3 - a)p + 9/4$. }

The Proposition 1 is proved in Appendix. 
It tells us about existence and uniqueness of
$R_{isco} > 2\mu$, which also obeys the inequality
 \begin{equation}
     R_{isco}  >  R_0,  
      \label{R_isco}
\end{equation}                                        %         
where $R_0$ is the radius of photon sphere, see (\ref{eq:r_0}).
 
{\bf Proposition 2.}
{\it The circular time-like geodesics for massive neutral point-like particles are described by arbitrary radius $R > R_0$, where $R_0$ is the radius of photonic sphere. The solutions with $R > R_{isco} = x_{isco} 2 \mu$ are stable, while solutions with $ R_0 < R < R_{isco}$ are unstable.} 
  
{\bf Proof.} It follows from Proposition 1  and relation ~\eqref{eq:W} that fourth order polynomial equation $W(R) = 0$ has one and only one root $R = R_{isco} $ for  $R   > 2 \mu $ which obeys $R_{isco} > R_0$, where $R_0$ is the radius of photon sphere (see (\ref{eq:r_0})). Hence $W(R) > 0$, or equivalently, $\frac{\partial^2 V}{\partial R^2} > 0$ for $R > R_{isco}$, i.e. the solutions are stable in this interval of $R$.   (See Eq. ~\eqref{d2VdR2}.)  For $R_0 < R < R_{isco}$ we have  $W(R) < 0$, or equivalently,  $\frac{\partial^2 V}{\partial R^2} < 0$, that means  that solutions are unstable for these values of $R$. 

\begin{figure}[ht]
\includegraphics[width=\columnwidth]{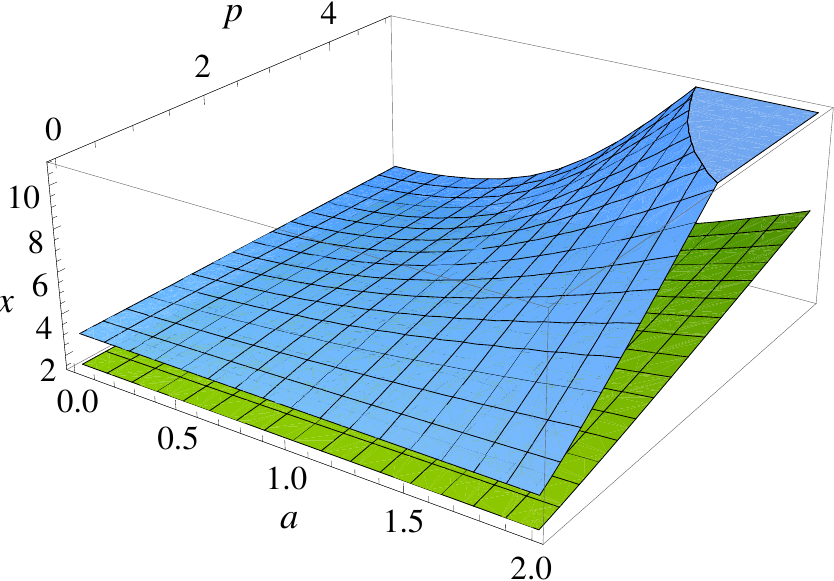}
\caption{Three dimensional plot of $x_{ISCO}$ and $x_0$ as functions of parameters $a$ and $p$.}
\label{fig:xiscox0}
\end{figure}

In Fig.~ \ref{fig:xiscox0} we constructed 3D plot of $x_{isco}=R_{isco}/(2\mu)$ and $x_0=R_{0}/(2\mu)$ as functions of $a$ and $p$. As one can observe $x_{isco}$ is always located above $x_0$, which proves the correctness our results.

\begin{figure}[ht]
\includegraphics[width=\columnwidth]{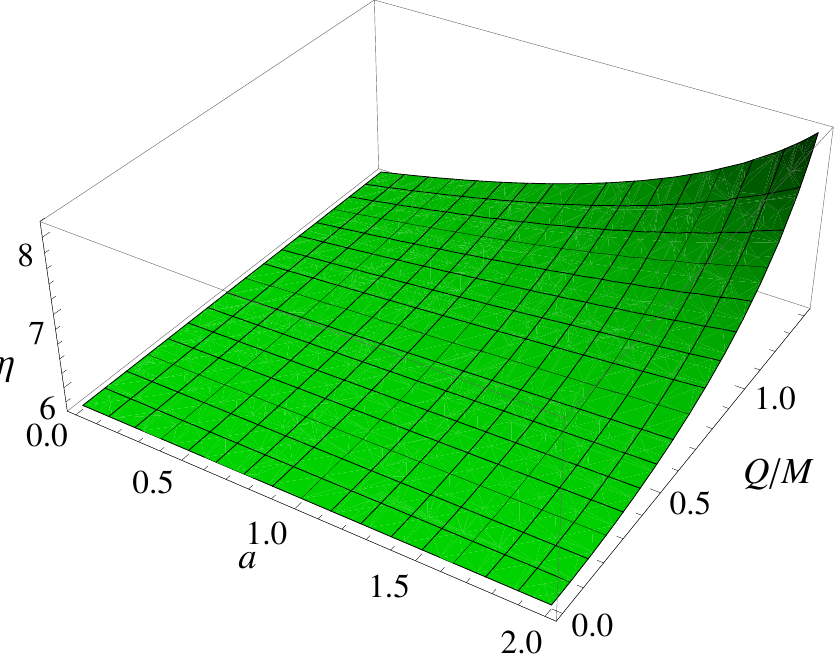}
\caption{Three dimensional plot of efficiency in \% as functions of parameters $a$ and $Q/M$.}
\label{fig:effi}
\end{figure}

In Fig.~ \ref{fig:effi} we constructed 3D plot of the efficiency converting matter into radiation and as functions of $a$ and $Q/M$. As one can observe the efficiency increases with increasing parameter $a$.

{\bf The case of big $p$.} It may readily  verified that for   $1 < a < 2$ we have the following asymptotic relation               
\begin{equation}
x_{isco}(a,p) \sim h(a) p,
\label{x1a2p_inf}
\end{equation}
as $ p \to +\infty $,  where
\begin{equation}
h(a) = (3/2)(a-1) + (1/2) \sqrt{(a-1)(5a-1)} > 0.
\label{h1a2}
\end{equation}
Indeed,   plugging the asymptotic relation $x = h p$, $h > 0$, into the master equation  (\ref{master}) and keeping the leading  terms  of  order $p^5$ we get a quadratic equation for $h$  
\begin{equation}
  h^2+ 3 (1-a)  h + (a-2)(a-1) = 0 ,
\label{eq_h_a}
\end{equation}
which gives us the positive solution (\ref{h1a2}) .
For $1< a < 2$ the relation (\ref{xbx0}) implies   
\begin{equation}
x_{0}(a,p) \sim (a - 1) p,
\end{equation}
as $ p \to +\infty $. It may be readily verified that inequality $x_0 < x_{isco}$ is satisfied for asymptotical values as $ p \to +\infty $.  

For $0 < a < 1$ we have another asymptotical relation
\begin{equation}
x_{isco}(a,p)  \to  x_{\infty}(a),
\label{x0a1p_inf}
\end{equation}
as $ p \to +\infty $,  where
\begin{equation}
x_{\infty}(a) =  \frac{\sqrt{-7a^2+10a+1}-4a^2+7a-1}{4a(1-a)} > 1 .
\label{x_inf_aless1}
\end{equation}
This relation may be obtained if the another asymptotic relation $x = const > 0$ will be substituted into the master eqution   (\ref{master}). Keeping the leading  terms  of  order $p^3$ we are led to another quadratic equation in $x$ 
\begin{equation}
        2(a-1)ax^2 - (4a^2-7a+1)x  + (a-1)(2a-3) = 0, 
\label{eq_x_a}
\end{equation}
which gives us a positive solution $x = x_{\infty}(a) > 0$ from  (\ref{x_inf_aless1}).
For $0< a < 1$ the relation (\ref{xbx0}) implies   
\begin{equation}
x_{0}(a,p) \to  \frac{3 -2a}{2(1 -a)},
\label{x01a2p_inf}
\end{equation}
as $ p \to +\infty $. It may be also  verified that inequality $x_0 < x_{isco}$ is satisfied in the limit $ p \to +\infty $.  

Let us fix the horizon radius $2 \mu$. Then, in the limit $P \to + \infty$, when $M \to + \infty$ and $|Q| \to +\infty$ (with saturation of inequality (\ref{i.18QM})) we obtain
\begin{equation}
R_{ISCO} \sim h(a) P \sim  h(a) |Q|/ \sqrt{2},
\label{R1a2P_inf}
\end{equation}
for $1< a < 2$ and 
\begin{equation}
R_{ISCO}  \to  x_{\infty}(a) 2 \mu,
%\label{x0a1p_inf}
\end{equation}
for $0 < a < 1$.
 
 %%%%%%%%%%%%%%%%%%%%%%%%%%%%%%%%%%%%%%%%%%%%%%%%%%%%%%%%%%%%%%%%%%%%%%%%%%%%%%%%%%%%%%%%%%%%%%%%%%%%%%%%%%%%%%%%%%%%%%%%%%%%%%%%%%%%%%
 
\section{Some Examples}\label{sec:examples}

For selected values of $a=0, 0.25, 1, 1.5, 2$, we overview the following expressions for $x_{isco}$.

For the Schwarzschild BH ($a=0$), the innermost stable circular orbit (ISCO) radius is:
\be \label{eq:xisco_0}
    x_{isco}(a=0)=3.
\ee

For $a=1/4$, the dimensionless ISCO radius is 
\bea \label{eq:xisco_025}
x_{isco}&=& \frac{1}{16} (12 - 9 p) 
+ \frac{\sqrt{X_{1/4}}}{32 (p + 4)}  \nonumber\\
&+& \frac{1}{2} \sqrt{\frac{A_{1/4}}{\sqrt{X_{1/4}}} - \frac{X_{1/4}}{256 (p + 4)^2} + T_{1/4}}, \quad \\\nonumber
X_{1/4}&=&256 (p +4)^2 Y_{1/4}^{\frac{1}{3}} \\\nonumber
&+& 4(p+4)(25p^3+92p^2+560p+576) \\\nonumber
&+& p^2(49p^4+364p^3+1148p^2+1216p+384) Y_{1/4}^{-\frac{1}{3}},\\\nonumber
Y_{1/4}&=& \frac{p^3 ( p +1) \sqrt{R_{1/4}}}{1024 (p + 4)^2} + Z_{1/4},\\\nonumber
\eea
\bea
R_{1/4}&=& (4 + p)^{-1}(4 +  7 p) \Big[16807p^6+247940p^5  \\ \nonumber
&+&1433180p^4 +  4978752p^3 +8065536p^2  \\
 &+& 5705728p+1417216 \Big],\nonumber
\\\nonumber
Z_{1/4} &=& 2^{-12} (p + 4)^{-3} p^3 \times \\ \nonumber
&\times&\Big[343p^6+3822p^5+21896p^4+76688p^3 \\ \nonumber
   &+&122016p^2+84224p+20480 \Big],\\ \nonumber
\eea
\bea 
\nonumber
A_{1/4}&=&\frac{1}{16}(27p^4 + 104p^3 + 2816p^2 + 9216p + 6912),\\\nonumber
T_{1/4}&=& \frac{3 (25p^3+92p^2+560 p+576)}{64 (4 + p)},\nonumber
\eea
where $p=P/2\mu$.

For  $a=1$, which corresponds to the Sen BH solution \cite{1992PhRvL..69.1006S},
the result is readily obtained \cite{BSIU}
\be \label{eq:xisco_1}
x_{isco}(a=1) = 1 + (1 + p)^{\frac{1}{3}} + (1 + p)^{\frac{2}{3}}.
\ee

For $a=2$, we are led to the Reissner--Nordstr\"{o}m BH case with the ISCO radius given by \cite{BSIU}
\be \label{eq:xisco_2}
  x_{isco}(a=2)= 1 + p + X_2^{\frac{1}{3}} + \frac{1 + p + p^2}{X_2^{\frac{1}{3}}},
\ee  
where
\be
X_2 = \frac{2 + p (1 + p) \Big[7 + 4 p (1 + p) + \sqrt{5 + 4 p (1 + p)}\Big]}{2 (1 + 2 p)}.\nonumber
\ee

Bounds for $Q/M$ are given by \eqref{i.18QM}. For the case $a=2$ we obtain $-\sqrt{2}<Q/M<\sqrt{2}$.
Reference \cite{MBI} demonstrates that, for the case $a=2$, applying the radial coordinate transformation  $R=r_{RN}-P$, along with $M_{RN}=\mu+P$, the metric \eqref{i4.9} becomes equivalent to the Reissner--Nordstr\"{o}m metric.

As expected, in the limiting case $Q\rightarrow 0$, $R_{ISCO}$ corresponds to the Schwarzschild value $6M$. In the cases of $0<a\leq 2$ the values of $R_{ISCO}$ depend upon the ratio of $Q/\mu$. 

Another key quantity of interest is the efficiency of matter-to-radiation conversion (see Ref. \cite{1973grav.book.....M}, p. 662 for details)
\begin{equation}
    \eta=[1-\tilde{E}(R_{ISCO})]\times100\%.
\end{equation}

Here $\tilde{E}(R_{ISCO}) = E(R_{ISCO})/m$ is given by relation \eqref{eq:E_of_R}  and 
$R_{ISCO} = 2 \mu x_4$, where  $x_4 = x_4 (a, p) > x_0 > 1$ is solution to master equation 
presented by a formula  \eqref{B_x4} from Appendix B. 

Figure ~\ref{fig:effi} shows the efficiency as a function of $Q/M$ and $a$.  As expected, the efficiency for various values of $a= 0.25, 0.5, 0.75, 1, 1.25, 1.5, 1.75, 2$, consistently exceeds that of the $a=0$ case \cite{BSIU}.

Generally, circular orbits are possible in the region $R_0<R<R_{ISCO}$; however, all such orbits are unstable.  Stable circular orbits require the following condition:

\be \label{eq:cond_of_stability}
   \frac{\partial^2V}{\partial R^2} > 0,
\ee
which is valid in the region $R > R_{ISCO}$.

%%%%%%%%%%%%%%%%%%%%%%%%%%%%%%%%%%%%%%%%%%%%%%%%%%%%%%%%%%%%%%%%%%%%%%%%%%%%%%%%%%%%%%%%%%%%%%%%%%%%%%%%%%%%%%%%%%%%
\section{Conclusion} \label{sec:conclusion}
%%%%%%%%%%%%%%%%%%%%%%%%%%%%%%%%%%%%%%%%%%%%%%%%%%%%%%%%%%%%%%%%%%%%%%%%%%%%%%%%%%%%%%%%%%%%%%%%%%%%%%%%%%%%%%%%%%%%

In this paper, we have explored the solution for a dyonic-like dilatonic BH. We have investigated circular geodesics of massive, neutral test particles and photons, by using various values for the metric parameters: $a$ , $\mu$ and $P$, which are related to the physical quantities: BH mass $M$ and the net charge $Q$. We have calculated $R_{ISCO}$, which is the radius of the innermost stable circular orbit for test particles in terms of these parameters. 
 
It satisfies the inequality $R_{ISCO} > R_{0}$, where $R_0$ is the radius of photonic sphere. The circular orbits of massive particles with radius $R$  which obey $R > R_{ISCO}$ are shown to be  stable  while those satisfying $R_0 < R < R_{ISCO}$ are shown  to be unstable. 

For all values of $a$ ($0 < a < 2$) the radius of ISCO is found by means of reducing the problem to the solution of fourth order polynomial equation $F(x) = 0$ - the so-called master equation for a dimensionless parameter $x$. The solution to master equation $x = x_{isco}$ which corresponds to ISCO radius is presented in Appendix ($x_{isco} = x_{4}$).

We have presented few explicit examples of solutions to master equation for   $a= 0, 1/4, 1, 2$ which contain a new one for  $a= 1/4$.  (The dimensionless parameter $a$ describes scalar products of dilatonic coupling vectors.)

The  analysis of stability of circular orbits  was  based  on examining the sign of the second derivative of the effective potential in the stationary point. Hopefully this second derivative has turned to be proportional to the fourth order polynomial $F(x)$ and this fact has drastically simplified our analysis of stability.

Here we have also found an analytical relation for the efficiency of converting matter into radiation by using the solution to master equation $x_{isco} = x_{4}$. The numerical analysis of the efficiency tells us that the  efficiency is larger for the case of Reissner--Nordstr\"{o}m BH ($a = 2$) $\eta=8.14\%$ and smaller for the case of Schwarzschild one ($a = 0$) $\eta=5.72\%$ \cite{2020PhRvD.102l4078B}. All other configurations lie within the parameter space bounded by these two cases. This peculiarity can be used to distinguish ordinary or astrophysical BHs from the dyonic-like dilatonic BHs. 

For a future research it looks reasonable to generalize  our setup by exploring the motion of a massive  point-like particle which carries two electric color charges:  $q_1$ and $q_2$, which correspond to  gauge subgroups: $(U(1))_1$ and  $(U(1))_2$, respectively.  

\begin{acknowledgements}
AM, GT, and GN are supported by Grant No. AP26194257; KB is supported by Grant No. AP19680128; and AU is supported by Grant No. BR21881941, all from the Science Committee of the Ministry of Science and Higher Education of the Republic of Kazakhstan. For VI the research was funded by RUDN University, scientific project number FSSF-2023-0003.
\end{acknowledgements}

\appendix

\section{}%{Appendix A}

Here we outline the proof of the Proposition 1.

{\bf Proof.} The existence of the root  satisfying (\ref{xb1}) can be readily proved.
Indeed,  due to  (\ref{master}) and (\ref{ap})  we have
\begin{equation}
F(1) = 2(p+1)^2((a - 2)p-2) < 0. 
     \label{F1_l0}            
\end{equation}                
Since   $(2ap+2) > 0$   we get     
  \begin{equation}
   F(x)    \to  + \infty ,      
    \label{Fxto_inf}  
  \end{equation}
 as $x \to  + \infty$. 
 This implies the existence of $x_{+} > 1$ such that 
  \begin{equation}
   F(x_{+}) > 1. 
    \label{Fx_p_b1}                                 \end{equation}
Applying the intermediate value theorem to the continuous function  $F(x)$  on closed interval $[1, x_{+}]$  with boundary values given by (\ref{F1_l0})  and  (\ref{Fx_p_b1}) we find that there exists   $x_{*}$  belonging to interval $(1, x_{+} )$ such that   
       \begin{equation}
       F(x_{*}  ) = 0.     
       \label{Fx_star_0} 
       \end{equation}
So, the existence of the solution to master equation (\ref{master}),  satisfying $x_{*}  > 1$, is proved. 
(Analogous consideration gives us an existence of another root $x_{**}$ ( $F(x_{**}  ) = 0$ ) which satisfies $x_{**}  < 1$.)
Now we prove the uniquenes of the root $x_{*}  > 1$. 
Let us suppose that there exists another root $x_{2,*}  > 1$.  Without loss of generality we put
 $x_{2,*} > x_{*}$. An elementary graphical analysis   leads us to three possibilities for our  quartic polynomial  $F(x)$:
 $$ i) \	F'(x_{*})  = 0,  \quad   F'(x_{2,*})  > 0, $$
 $$ii) \	F'(x_{*})  > 0,   \quad   F'(x_{2,*})  = 0,$$
 $$iii) \	F'(x_{*})  > 0,  \quad    F'(x_{2,*})  < 0,  
 \quad F'(x_{3,*})  > 0, \quad $$     
where   $x_{3,*}$ is the third root: $F(x_{3,*})  = 0$,  obeying (without loss of generality) $x_{3,*} > x_{2,*} $. 
 
In any case by Rolle's theorem we get that there exist two stationary points of the function $F(x)$ : $ 1 < x_{1} < x_{2}$, which obey the cubic equation     $$F'(x_{i} ) = 0,$$   $i = 1,2$. But according to the Lemma proved below   this is impossible. 
 
Thus, we come to a contradiction which proves the uniqueness of the root $x_{*} > 1$ of our quartic equation. 

Now let us prove that $x_{*} > x_0$, where $x_0$ is defined in equation (4) of the question.

Let us consider the auxiliary function $$v(x)  = \frac{H^a x^2 \left[ap(x-1) + p+x\right]}{2 \Delta_0},$$ where $H = 1 + p/x$, and quadratic polynomial $$\Delta_0 =  x^2 + ((1 - a)p - 3/2)x + (2a-3) p/2$$ has a biggest real root $x_0 > 1$. The function $v(x)$ is well defined on interval $(x_0, + \infty)$ and tends to $+ \infty$ in two limits: when $x \to x_0$ or $x \to + \infty$. Thus, it has global minimum at some point $x_{0,*}$. We get that $v'(x_{0,*}) = 0$. But due to identity (which could be readily verified) 
$$ \frac{dv}{dx} = \frac{H^{a-1}}{4 \Delta_0^2}  F,$$ 
we get that $F(x_{0,*}) = 0$. (Here we have used in fact  relations (\ref{eq:dL_of_dR}) and (\ref{eq:W}) rewritten in dimensionless form.)

Thus, there exists a root $x_{0,*}$ of our quartic equation which obeys    $1< x_0 < x_{0,*}$. But due to uniqueness of a root obeying $x > 1$ we get $x_{0,*} = x_{*}$. Thus, $1< x_0 < x_{*}$. Hence Proposition 1  is  proved.   

{\bf Lemma.} Let us consider the  function 
  \begin{eqnarray}
 F'(x) = 4(2ap+2)x^3 \qquad  \qquad \label{L1} \\ 
 +3(6(1-a)ap^2+(6-12a)p-6)x^2  \qquad 
 \nonumber \\ 
     +2p(2(a-2)(a-1)ap^2 \qquad 
     \nonumber \\ 
     +3(5a^2-9a+2)p+12a-18)x  \qquad 
    \nonumber \\  
    - p^2((a-2)(4a^2-7a+1)p+9a^2-25a+18). \quad 
     \nonumber
\end{eqnarray}                                
Here $0 < a < 2$ and $p > 0$. Then, the function $F'(x)$ has no more than one root in the interval $(1, + \infty)$. Moreover, it has one root if 
 $$(a) \  F'(1) = (p+1) \times $$
 $$ \times [(a-2)(3a-1)p^2 - 4(a+2)p -10] < 0 $$
     and no roots if  
  $$(b)  \qquad F'(1) \geq 0. $$

{\bf Proof}. First, we consider the  case (b), i. e. when 
 \begin{equation}
  (2 - a)(3a-1)p^2 + 4(a+2)p + 10  \leq 0.  \label{L2}
  \end{equation}
 
This relation implies $0 < a < 1/3$ since otherwise (for  $1/3 \leq a < 2$) the left hand side in  \eqref{L2} is positive.  Due to \eqref{L2} or 
\begin{equation}         
(2 - a)(1 - 3a)p^2 - 4(a+2)p - 10  \geq 0. 
\label{L3a} 
\end{equation}
we get 
\begin{equation}
p \geq  p(a) \equiv \frac{4(a +2) + \sqrt{16(a+2)^2 + 40(2 - a)(1 - 3a)}}{2(1-3a)(2-a)}   \label{L4}        \end{equation}                    
The function $p(a)$ is monotonically increasing in interval  $(0,1/3)$ since 
$$ \frac{ \sqrt{16(a+2)^2 + 40(2 - a)(1 - 3a)}}{(1-3a)} = $$ $$
\sqrt{16(a+2)^2 (1-3a)^{-2}  + 40(2 - a)(1 - 3a)^{-1}} $$
is monotonically increasing in  $a \in (0,1/3)$. 
We note that due to the limit $p(+0) = 5$ we obtain
\begin{equation}
  p(a) > 5, \qquad \label{L5a} 
\end{equation} 
for all $a$ belonging to $(0,1/3)$.  
Now we prove that in case (b) 
\begin{eqnarray}
& F''(x) = 12(2ap+2)x^2 \label{L5}  \\ 
 +&6(6(1-a)ap^2+(6-12a)p-6)x+ 
 \nonumber \\  
 &+2p[2(a-2)(a-1)ap^2
 \nonumber \\ 
 &+3(5a^2-9a+2)p+12a-18] > 0  
 \quad \nonumber
 \end{eqnarray} 
for all $x > 1$, $0 < a < 1/3$ and $p \geq p(a)$,  implying
 \begin{equation}
  F'(x) > 0  \quad   \label{L6}   
 \end{equation}   
for all $x > 1$, $0 < a < 1/3$, $p \geq p(a)$, which proves
 ``one half'' of the Lemma for the case (b). 

In order to prove inequality 
\eqref{L5} it is sufficient to prove two inequalities
\begin{equation}
V = 6(1-a)ap^2+(6-12a)p-6 > 0, \qquad  \label{L6V}
\end{equation}
     and
\begin{equation}
Z = 2(a-2)(a-1)ap^2+3(5a^2-9a+2)p+12a-18 > 0  
\label{L6Z}
\end{equation}
for all $x > 1$, $0 < a < 1/3$, $p \geq p(a)$. 
 Plugging the  bound 
   $$p^2 \geq \frac{(4(a+2)p + 10)}{(2 - a)(1 - 3a)},$$
following from  \eqref{L3a}, into relation for $V$ 
we obtain 
\begin{eqnarray}
 V \geq V_b \equiv  \frac{6(1-a)a (4(a+2)p + 10)}{(2 - a)(1 - 3a)} + (6-12a)p-6  \nonumber \\
=   \frac{6(-10a^3p+13a^2p-3ap+2p-13a^2+17a-2)}{(2-a)(1 -3a)},
\nonumber \\ 
   \label{L7V} 
\end{eqnarray}
  and  
  \begin{eqnarray}
Z \geq Z_b \equiv 
\frac{2(a-2)(a-1)a(4(a+2)p + 10)}{(2 - a)(1 - 3a)} 
\nonumber \\ 
+3(5a^2-9a+2)p  +12a-18  \nonumber \\ 
= \frac{-53a^3p+88a^2p-29ap+6p-56a^2+86a-18}{1-3a}. 
\nonumber \\ 
\label{L7Z}
\end{eqnarray}
In order to prove the inequalities \eqref{L6V} and 
\eqref{L6Z} one needs (due to \eqref{L7V} and \eqref{L7Z})
 to prove
 \begin{equation}
v = (-10a^3+13a^2-3a+2)p-13a^2+17a-2 > 0  \label{L8v} 
\end{equation}
and 
\begin{equation}
z = (-53a^3 + 88a^2 - 29a + 6)p - 56a^2 + 86a - 18 > 0, 
 \label{L8z}
\end{equation}
for all $0 < a < 1/3$ and $p \geq p(a)$. 
  
The first bound can be readily proved. Indeed,  for the bracket in \eqref{L8v} we have
\begin{equation}
(-10a^3+13a^2-3a+2) >  -10 \left(\frac{1}{3}\right)^3
 - 3 \left(\frac{1}{3}\right) +2 =\frac{17}{27}
 \label{L9av}
 \end{equation}
and hence (using (5a): $ p(a) > 5 $) we get
 \begin{equation}   
  v   > \frac{17}{27}5 - \frac{13}{9} - 2 = \frac{5}{27} > 0 
  \label{L9v}
 \end{equation} 
Now we prove the second bound \eqref{L8z}. For the bracket in  
\eqref{L8z} we use the following inequality 
\begin{equation}
 z_1(a) = (-53a^3 + 88a^2 - 29a + 6) >  3 > 0, 
 \label{L9z1} 
\end{equation}
for all $0 < a < 1/3$. This can be readily obtained from graphical analysis of the function  $z_1(a)$ on interval $(0,1/3)$. 
This function reaches a minimum at the point 
\begin{equation}
  a_{min} = -(\sqrt{3133}-88)/159 \approx 0.2014 \quad   \label{L10a}
\end{equation}
(which could be obtained from quadratic equation $\frac{dz_1(a)}{da} = 0$) 
with the value 
\begin{equation}
z_1(a_{min}) = \frac{600698 - 2(3133)^{3/2}}{75843}        
       \approx 3.296 >  3. \quad    \label{L10z1}
\end{equation}
Since  $z_1(a) > 0$, we obtain  for $p \geq p(a)$
\begin{eqnarray}
z = (-53a^3+88a^2-29a+6)p-56a^2+86a-18 
\nonumber \\ 
\geq z_b (a) = (-53a^3 + 88a^2-29a+6) p(a)
\nonumber \\    
-56a^2+86a-18 > 12. \qquad \label{L11z} 
\end{eqnarray}

The last bound \eqref{L11z} just follows from the graphical analysis of the function 
$z_b (a)$ on the interval (0, 1/3). This function is monotonically increasing  from $(12_{+0})$ to $+ \infty$.  Thus, relation \eqref{L8z} is proved and hence 
the part (b) of the Lemma is also proved.

Now we consider the  case (a) $F'(1) < 0$. Here we get two subcases 
  \begin{equation}
      (a1)  \quad  1/3 \leq  a < 2, \quad  p > 0,    
      \label{L12a1}
  \end{equation}
  and
  \begin{equation}
(a2)   \  0 < a < 1/3,  \qquad   0 < p < p(a), 
      \label{L12a2}
 \end{equation}     
where   $ p(a)$ is defined in (4).   
Since  $F'(1) < 0$ and $F'(+\infty) = + \infty$,  then due to Intermediate Value Theorem  for all $a$  and  $p$ from  \eqref{L12a1} and \eqref{L12a2} there exists at least one root $x_{*}$ of the cubic polynomial  $F'(x)$  in the interval  $(1, + \infty)$, i.e. 
  \begin{equation}
   F'(x_{*})  = 0,   \qquad  x_{*} > 1.   \label{L13} 
  \end{equation}
          
 Let us suppose that for some $a$ and $p$ there exists another root   $x_{**} \neq x_{*}$,   i.e. $F'(x_{**})  = 0$. Without loss of generality we put    $x_{*}  < x_{**}$.  
 An elementary graphical analysis leads us to three variants for our  cubic polynomial  $F'(x)$:
 $$ i) \	F''(x_{*})  = 0,  \quad   F''(x_{**})  > 0, $$
$$ii) \	F''(x_{*})  > 0,   \quad   F''(x_{**})  = 0,$$
$$iii) \	F''(x_{*})  > 0,  \quad    F''(x_{**})  < 0,  
\quad F''(x_{***})  > 0, $$     
where   $x_{***}$ is the third root: $F'(x_{***})  = 0$, obeying (without loss of generality) $x_{***} > x_{**} $.  
By Rolle's(-Cauchy's) theorem we find that in all three cases there exist two different roots $x_1$, $x_2$   of  quadratic polynomial 
 \begin{eqnarray}
   H = F''(x) = 12(2ap+2)x^2 \qquad   \nonumber \\
      +6(6(1-a)ap^2  +(6-12a)p-6)x \qquad   \nonumber \\
      +  2p(2(a-2)(a-1)ap^2 
   \qquad     \nonumber \\
   +3(5a^2-9a+2)p+12a-18)  = 0,  
    \label{L15}
  \end{eqnarray} 
which satisfy $x_1 \neq x_2$ and
\begin{equation}
    F''(x_1) = F''(x_2) = 0. \qquad  \label{L16} 
\end{equation}
By making the redefinition of our variable: $y = x - 1$, we rewrite the  quadratic equation  \eqref{L15} as 
\begin{eqnarray}
 H = H(y) = 24(ap+1)y^2+  \qquad  \nonumber \\ 
 12(3a(1-a)p^2+(3-2a)p+1)y  \qquad 
 \nonumber \\ 
+(4(a - 1)(a - 2)ap^3 \qquad 
\nonumber \\ 
+ 6(-a^2- 3a+2)p^2 -24ap-12 = 0.  \qquad 
 \label{L17}
 \end{eqnarray} 
This quadratic polynomial $H(y)$ should have two different  positive roots $y_1 = x_1 - 1 >0$ and $y_2 = x_2 - 1 > 0$. Due to Vieta's formulas and  $ ap+1 >0$ 
we obtain the following inequalities:
\begin{eqnarray}
  3a(1-a)p^2+(3-2a)p+1 <  0,  \qquad   \label{L18B} \\
  4(a - 1)(a - 2)ap^3 \qquad
  \nonumber \\ 
  + 6(-a^2- 3a+2)p^2 -24ap-12 > 0.  \label{L18C}
\end{eqnarray} 
   Due to \eqref{L18B}  we get
   \begin{equation}
    1 < a < 2. \qquad     \label{L19}
   \end{equation} 
It can be readily seen that for this values of  $a$  all coefficients of the cubic polynomial in the left hand side of  \eqref{L18C} are negative and  hence the inequality \eqref{L18C} is not satisfied for all $p > 0$ and $1 < a < 2$. 
Thus,  in case (a)  we have only one root of cubic polynomial $F'(x)$ belonging to $(1, + \infty)$. This finishes the proof of the Lemma. 

\section{}%{Appendix B}

In this Appendix we show the solution of the master equation using Mathematica software. For convenience, we use different notations from the previous Appendix. So, here the master equation is given by:

\begin{equation}
A_4 x^4 + A_3 x^3 + A_2 x^2 + A_1 x + A_0 = 0 ,  \end{equation}
where the coefficients $A_i$ are
\begin{eqnarray}
   A_0 &=& (a - 2)(a - 1)(2a - 3)p^3 ,\\ 
   A_1 &=& (-18 + 25a - 9a^2)p^2 \\ \nonumber
   &+& (2 - a)(1 - 7a + 4a^2)p^3 ,\\
   A_2 &=& 6(2a - 3)p + 3(2 - 9a + 5a^2)p^2 ,\\   \nonumber
   &+& 2a(a - 1)(a - 2)p^3\\
   A_3 &=& -6 + 6(1 - 2a)p + 6a(1 - a)p^2 ,\\
   A_4 &=& 2(1 + ap).
\end{eqnarray}

The solutions of the master equation are shown as
\begin{eqnarray}
x_1 &=& -\frac{A_3}{4A_4} - B_7 - B_8 ,
\label{B_x1} \\
x_2 &=& -\frac{A_3}{4A_4} - B_7 + B_8 ,
\label{B_x2} 
\end{eqnarray}
\begin{eqnarray}
x_3 &=& -\frac{A_3}{4A_4} + B_7 - B_9 ,
\label{B_x3} \\
x_4 &=& -\frac{A_3}{4A_4} + B_7 + B_9 . \label{B_x4}
\end{eqnarray}

Only $x_4$ corresponds to physical ISCO solution and other solutions are either essentially complex or negative or smaller than the dimensionless photon sphere radius.

The remaining coefficients are listed as follows:
\begin{eqnarray}
   B_9 &=& \frac{1}{2} \sqrt{ B_2 - \frac{2^{1/3} B_0}{3 A_4 B_6} - \frac{B_6}{3 \cdot 2^{1/3} A_4} + \frac{B_3}{8 B_7}} ,\\
   B_8 &=& \frac{1}{2} \sqrt{ B_2 - \frac{2^{1/3} B_0}{3 A_4 B_6} - \frac{B_6}{3 \cdot 2^{1/3} A_4} - \frac{B_3}{8 B_7}} ,\\
   B_7 &=& \frac{1}{2} \sqrt{ \frac{B_2}{2} + \frac{2^{1/3} B_0}{3 A_4 B_6} + \frac{B_6}{3 \cdot 2^{1/3} A_4}} ,   
\end{eqnarray}
\begin{eqnarray}
    B_6 &=& (B_4 + B_5)^{1/3} ,\\
    B_5 &=& \sqrt{ B_4^2 - 4 B_0^3} ,\\
    B_4 &=& B_1-27 A_0 A_3^2 + 72 A_0 A_2 A_4 + 54 A_0 A_4^2 B_2 ,\quad
\end{eqnarray}
\begin{eqnarray}
    B_3 &=& -\frac{A_3^3}{A_4^3} + \frac{4 A_2 A_3}{A_4^2} - \frac{8 A_1}{A_4} ,\\
    B_2 &=& \frac{A_3^2}{2 A_4^2} - \frac{4 A_2}{3 A_4} ,\\
    B_1 &=& 2 A_2^3 - 9 A_2 (A_1 A_3 + 8 A_0 A_4)\\ \nonumber &+& 27 (A_0 A_3^2 + A_1^2 A_4) ,\\
    B_0 &=& A_2^2 - 3 A_1 A_3 + 12 A_0 A_4 .
\end{eqnarray}


\begin{thebibliography}{61}%
\makeatletter
\providecommand \@ifxundefined [1]{%
 \@ifx{#1\undefined}
}%
\providecommand \@ifnum [1]{%
 \ifnum #1\expandafter \@firstoftwo
 \else \expandafter \@secondoftwo
 \fi
}%
\providecommand \@ifx [1]{%
 \ifx #1\expandafter \@firstoftwo
 \else \expandafter \@secondoftwo
 \fi
}%
\providecommand \natexlab [1]{#1}%
\providecommand \enquote  [1]{``#1''}%
\providecommand \bibnamefont  [1]{#1}%
\providecommand \bibfnamefont [1]{#1}%
\providecommand \citenamefont [1]{#1}%
\providecommand \href@noop [0]{\@secondoftwo}%
\providecommand \href [0]{\begingroup \@sanitize@url \@href}%
\providecommand \@href[1]{\@@startlink{#1}\@@href}%
\providecommand \@@href[1]{\endgroup#1\@@endlink}%
\providecommand \@sanitize@url [0]{\catcode `\\12\catcode `\$12\catcode `\&12\catcode `\#12\catcode `\^12\catcode `\_12\catcode `\%12\relax}%
\providecommand \@@startlink[1]{}%
\providecommand \@@endlink[0]{}%
\providecommand \url  [0]{\begingroup\@sanitize@url \@url }%
\providecommand \@url [1]{\endgroup\@href {#1}{\urlprefix }}%
\providecommand \urlprefix  [0]{URL }%
\providecommand \Eprint [0]{\href }%
\providecommand \doibase [0]{https://doi.org/}%
\providecommand \selectlanguage [0]{\@gobble}%
\providecommand \bibinfo  [0]{\@secondoftwo}%
\providecommand \bibfield  [0]{\@secondoftwo}%
\providecommand \translation [1]{[#1]}%
\providecommand \BibitemOpen [0]{}%
\providecommand \bibitemStop [0]{}%
\providecommand \bibitemNoStop [0]{.\EOS\space}%
\providecommand \EOS [0]{\spacefactor3000\relax}%
\providecommand \BibitemShut  [1]{\csname bibitem#1\endcsname}%
\let\auto@bib@innerbib\@empty
%</preamble>
\bibitem [{\citenamefont {{Abbott et al}}(2016{\natexlab{a}})}]{2016PhRvL.116f1102A}%
  \BibitemOpen
  \bibfield  {author} {\bibinfo {author} {\bibfnamefont {B.~P.}\ \bibnamefont {{Abbott et al}}},\ }\bibfield  {title} {\bibinfo {title} {{Observation of Gravitational Waves from a Binary Black Hole Merger}},\ }\href {https://doi.org/10.1103/PhysRevLett.116.061102} {\bibfield  {journal} {\bibinfo  {journal} {\prl}\ }\textbf {\bibinfo {volume} {116}},\ \bibinfo {eid} {061102} (\bibinfo {year} {2016}{\natexlab{a}})},\ \Eprint {https://arxiv.org/abs/1602.03837} {arXiv:1602.03837 [gr-qc]} \BibitemShut {NoStop}%
\bibitem [{\citenamefont {{Abbott et al}}(2016{\natexlab{b}})}]{2016PhRvL.116v1101A}%
  \BibitemOpen
  \bibfield  {author} {\bibinfo {author} {\bibfnamefont {B.~P.}\ \bibnamefont {{Abbott et al}}},\ }\bibfield  {title} {\bibinfo {title} {{Tests of General Relativity with GW150914}},\ }\href {https://doi.org/10.1103/PhysRevLett.116.221101} {\bibfield  {journal} {\bibinfo  {journal} {\prl}\ }\textbf {\bibinfo {volume} {116}},\ \bibinfo {eid} {221101} (\bibinfo {year} {2016}{\natexlab{b}})},\ \Eprint {https://arxiv.org/abs/1602.03841} {arXiv:1602.03841 [gr-qc]} \BibitemShut {NoStop}%
\bibitem [{\citenamefont {{Ghez}}\ \emph {et~al.}(1998)\citenamefont {{Ghez}}, \citenamefont {{Klein}}, \citenamefont {{Morris}},\ and\ \citenamefont {{Becklin}}}]{1998ApJ...509..678G}%
  \BibitemOpen
  \bibfield  {author} {\bibinfo {author} {\bibfnamefont {A.~M.}\ \bibnamefont {{Ghez}}}, \bibinfo {author} {\bibfnamefont {B.~L.}\ \bibnamefont {{Klein}}}, \bibinfo {author} {\bibfnamefont {M.}~\bibnamefont {{Morris}}},\ and\ \bibinfo {author} {\bibfnamefont {E.~E.}\ \bibnamefont {{Becklin}}},\ }\bibfield  {title} {\bibinfo {title} {{High Proper-Motion Stars in the Vicinity of Sagittarius A*: Evidence for a Supermassive Black Hole at the Center of Our Galaxy}},\ }\href {https://doi.org/10.1086/306528} {\bibfield  {journal} {\bibinfo  {journal} {\apj}\ }\textbf {\bibinfo {volume} {509}},\ \bibinfo {pages} {678} (\bibinfo {year} {1998})},\ \Eprint {https://arxiv.org/abs/astro-ph/9807210} {arXiv:astro-ph/9807210 [astro-ph]} \BibitemShut {NoStop}%
\bibitem [{\citenamefont {{Ghez}}\ \emph {et~al.}(2000)\citenamefont {{Ghez}}, \citenamefont {{Morris}}, \citenamefont {{Becklin}}, \citenamefont {{Tanner}},\ and\ \citenamefont {{Kremenek}}}]{2000Natur.407..349G}%
  \BibitemOpen
  \bibfield  {author} {\bibinfo {author} {\bibfnamefont {A.~M.}\ \bibnamefont {{Ghez}}}, \bibinfo {author} {\bibfnamefont {M.}~\bibnamefont {{Morris}}}, \bibinfo {author} {\bibfnamefont {E.~E.}\ \bibnamefont {{Becklin}}}, \bibinfo {author} {\bibfnamefont {A.}~\bibnamefont {{Tanner}}},\ and\ \bibinfo {author} {\bibfnamefont {T.}~\bibnamefont {{Kremenek}}},\ }\bibfield  {title} {\bibinfo {title} {{The accelerations of stars orbiting the Milky Way's central black hole}},\ }\href {https://doi.org/10.1038/35030032} {\bibfield  {journal} {\bibinfo  {journal} {\nat}\ }\textbf {\bibinfo {volume} {407}},\ \bibinfo {pages} {349} (\bibinfo {year} {2000})},\ \Eprint {https://arxiv.org/abs/astro-ph/0009339} {arXiv:astro-ph/0009339 [astro-ph]} \BibitemShut {NoStop}%
\bibitem [{\citenamefont {{Gravity Collaboration}}(2018)}]{2018A&A...615L..15G}%
  \BibitemOpen
  \bibfield  {author} {\bibinfo {author} {\bibnamefont {{Gravity Collaboration}}},\ }\bibfield  {title} {\bibinfo {title} {{Detection of the gravitational redshift in the orbit of the star S2 near the Galactic centre massive black hole}},\ }\href {https://doi.org/10.1051/0004-6361/201833718} {\bibfield  {journal} {\bibinfo  {journal} {\aap}\ }\textbf {\bibinfo {volume} {615}},\ \bibinfo {eid} {L15} (\bibinfo {year} {2018})},\ \Eprint {https://arxiv.org/abs/1807.09409} {arXiv:1807.09409 [astro-ph.GA]} \BibitemShut {NoStop}%
\bibitem [{\citenamefont {{Event Horizon Telescope Collaboration}}(2019)}]{2019ApJ...875L...1E}%
  \BibitemOpen
  \bibfield  {author} {\bibinfo {author} {\bibnamefont {{Event Horizon Telescope Collaboration}}},\ }\bibfield  {title} {\bibinfo {title} {{First M87 Event Horizon Telescope Results. I. The Shadow of the Supermassive Black Hole}},\ }\href {https://doi.org/10.3847/2041-8213/ab0ec7} {\bibfield  {journal} {\bibinfo  {journal} {\apjl}\ }\textbf {\bibinfo {volume} {875}},\ \bibinfo {eid} {L1} (\bibinfo {year} {2019})},\ \Eprint {https://arxiv.org/abs/1906.11238} {arXiv:1906.11238 [astro-ph.GA]} \BibitemShut {NoStop}%
\bibitem [{\citenamefont {{Event Horizon Telescope Collaboration}}(2022{\natexlab{a}})}]{2022ApJ...930L..12E}%
  \BibitemOpen
  \bibfield  {author} {\bibinfo {author} {\bibnamefont {{Event Horizon Telescope Collaboration}}},\ }\bibfield  {title} {\bibinfo {title} {{First Sagittarius A* Event Horizon Telescope Results. I. The Shadow of the Supermassive Black Hole in the Center of the Milky Way}},\ }\href {https://doi.org/10.3847/2041-8213/ac6674} {\bibfield  {journal} {\bibinfo  {journal} {\apjl}\ }\textbf {\bibinfo {volume} {930}},\ \bibinfo {eid} {L12} (\bibinfo {year} {2022}{\natexlab{a}})}\BibitemShut {NoStop}%
\bibitem [{\citenamefont {Sotiriou}(2006)}]{Sotiriou:2006hs}%
  \BibitemOpen
  \bibfield  {author} {\bibinfo {author} {\bibfnamefont {T.~P.}\ \bibnamefont {Sotiriou}},\ }\bibfield  {title} {\bibinfo {title} {{f(R) gravity and scalar-tensor theory}},\ }\href {https://doi.org/10.1088/0264-9381/23/17/003} {\bibfield  {journal} {\bibinfo  {journal} {Class. Quantum Gravity}\ }\textbf {\bibinfo {volume} {23}},\ \bibinfo {pages} {5117} (\bibinfo {year} {2006})},\ \Eprint {https://arxiv.org/abs/gr-qc/0604028} {arXiv:gr-qc/0604028} \BibitemShut {NoStop}%
\bibitem [{\citenamefont {{Clifton}}\ \emph {et~al.}(2012)\citenamefont {{Clifton}}, \citenamefont {{Ferreira}}, \citenamefont {{Padilla}},\ and\ \citenamefont {{Skordis}}}]{2012PhR...513....1C}%
  \BibitemOpen
  \bibfield  {author} {\bibinfo {author} {\bibfnamefont {T.}~\bibnamefont {{Clifton}}}, \bibinfo {author} {\bibfnamefont {P.~G.}\ \bibnamefont {{Ferreira}}}, \bibinfo {author} {\bibfnamefont {A.}~\bibnamefont {{Padilla}}},\ and\ \bibinfo {author} {\bibfnamefont {C.}~\bibnamefont {{Skordis}}},\ }\bibfield  {title} {\bibinfo {title} {{Modified gravity and cosmology}},\ }\href {https://doi.org/10.1016/j.physrep.2012.01.001} {\bibfield  {journal} {\bibinfo  {journal} {\physrep}\ }\textbf {\bibinfo {volume} {513}},\ \bibinfo {pages} {1} (\bibinfo {year} {2012})},\ \Eprint {https://arxiv.org/abs/1106.2476} {arXiv:1106.2476 [astro-ph.CO]} \BibitemShut {NoStop}%
\bibitem [{\citenamefont {Astashenok}\ \emph {et~al.}(2015)\citenamefont {Astashenok}, \citenamefont {Capozziello},\ and\ \citenamefont {Odintsov}}]{Astashenok:2014nua}%
  \BibitemOpen
  \bibfield  {author} {\bibinfo {author} {\bibfnamefont {A.~V.}\ \bibnamefont {Astashenok}}, \bibinfo {author} {\bibfnamefont {S.}~\bibnamefont {Capozziello}},\ and\ \bibinfo {author} {\bibfnamefont {S.~D.}\ \bibnamefont {Odintsov}},\ }\bibfield  {title} {\bibinfo {title} {{Extreme neutron stars from Extended Theories of Gravity}},\ }\href {https://doi.org/10.1088/1475-7516/2015/01/001} {\bibfield  {journal} {\bibinfo  {journal} {JCAP}\ }\textbf {\bibinfo {volume} {01}},\ \bibinfo {pages} {001}},\ \Eprint {https://arxiv.org/abs/1408.3856} {arXiv:1408.3856 [gr-qc]} \BibitemShut {NoStop}%
\bibitem [{\citenamefont {Astashenok}\ \emph {et~al.}(2013)\citenamefont {Astashenok}, \citenamefont {Capozziello},\ and\ \citenamefont {Odintsov}}]{Astashenok:2013vza}%
  \BibitemOpen
  \bibfield  {author} {\bibinfo {author} {\bibfnamefont {A.~V.}\ \bibnamefont {Astashenok}}, \bibinfo {author} {\bibfnamefont {S.}~\bibnamefont {Capozziello}},\ and\ \bibinfo {author} {\bibfnamefont {S.~D.}\ \bibnamefont {Odintsov}},\ }\bibfield  {title} {\bibinfo {title} {{Further stable neutron star models from f(R) gravity}},\ }\href {https://doi.org/10.1088/1475-7516/2013/12/040} {\bibfield  {journal} {\bibinfo  {journal} {JCAP}\ }\textbf {\bibinfo {volume} {12}},\ \bibinfo {pages} {040}},\ \Eprint {https://arxiv.org/abs/1309.1978} {arXiv:1309.1978 [gr-qc]} \BibitemShut {NoStop}%
\bibitem [{\citenamefont {Astashenok}\ \emph {et~al.}(2017)\citenamefont {Astashenok}, \citenamefont {Odintsov},\ and\ \citenamefont {de~la Cruz-Dombriz}}]{Astashenok:2017dpo}%
  \BibitemOpen
  \bibfield  {author} {\bibinfo {author} {\bibfnamefont {A.~V.}\ \bibnamefont {Astashenok}}, \bibinfo {author} {\bibfnamefont {S.~D.}\ \bibnamefont {Odintsov}},\ and\ \bibinfo {author} {\bibfnamefont {A.}~\bibnamefont {de~la Cruz-Dombriz}},\ }\bibfield  {title} {\bibinfo {title} {{The realistic models of relativistic stars in $f(R) = R + \alpha R^2$ gravity}},\ }\href {https://doi.org/10.1088/1361-6382/aa8971} {\bibfield  {journal} {\bibinfo  {journal} {Class. Quantum Gravity}\ }\textbf {\bibinfo {volume} {34}},\ \bibinfo {pages} {205008} (\bibinfo {year} {2017})},\ \Eprint {https://arxiv.org/abs/1704.08311} {arXiv:1704.08311 [gr-qc]} \BibitemShut {NoStop}%
\bibitem [{\citenamefont {Capozziello}\ \emph {et~al.}(2019)\citenamefont {Capozziello}, \citenamefont {D'Agostino},\ and\ \citenamefont {Luongo}}]{Capozziello:2019cav}%
  \BibitemOpen
  \bibfield  {author} {\bibinfo {author} {\bibfnamefont {S.}~\bibnamefont {Capozziello}}, \bibinfo {author} {\bibfnamefont {R.}~\bibnamefont {D'Agostino}},\ and\ \bibinfo {author} {\bibfnamefont {O.}~\bibnamefont {Luongo}},\ }\bibfield  {title} {\bibinfo {title} {{Extended Gravity Cosmography}},\ }\href {https://doi.org/10.1142/S0218271819300167} {\bibfield  {journal} {\bibinfo  {journal} {Int. J. Mod. Phys. D}\ }\textbf {\bibinfo {volume} {28}},\ \bibinfo {pages} {1930016} (\bibinfo {year} {2019})},\ \Eprint {https://arxiv.org/abs/1904.01427} {arXiv:1904.01427 [gr-qc]} \BibitemShut {NoStop}%
\bibitem [{\citenamefont {Astashenok}\ \emph {et~al.}(2020)\citenamefont {Astashenok}, \citenamefont {Capozziello}, \citenamefont {Odintsov},\ and\ \citenamefont {Oikonomou}}]{Astashenok:2020qds}%
  \BibitemOpen
  \bibfield  {author} {\bibinfo {author} {\bibfnamefont {A.~V.}\ \bibnamefont {Astashenok}}, \bibinfo {author} {\bibfnamefont {S.}~\bibnamefont {Capozziello}}, \bibinfo {author} {\bibfnamefont {S.~D.}\ \bibnamefont {Odintsov}},\ and\ \bibinfo {author} {\bibfnamefont {V.~K.}\ \bibnamefont {Oikonomou}},\ }\bibfield  {title} {\bibinfo {title} {{Extended Gravity Description for the GW190814 Supermassive Neutron Star}},\ }\href {https://doi.org/10.1016/j.physletb.2020.135910} {\bibfield  {journal} {\bibinfo  {journal} {Phys. Lett. B}\ }\textbf {\bibinfo {volume} {811}},\ \bibinfo {pages} {135910} (\bibinfo {year} {2020})},\ \Eprint {https://arxiv.org/abs/2008.10884} {arXiv:2008.10884 [gr-qc]} \BibitemShut {NoStop}%
\bibitem [{\citenamefont {{Event Horizon Telescope Collaboration}}(2022{\natexlab{b}})}]{2022ApJ...930L..17E}%
  \BibitemOpen
  \bibfield  {author} {\bibinfo {author} {\bibnamefont {{Event Horizon Telescope Collaboration}}},\ }\bibfield  {title} {\bibinfo {title} {{First Sagittarius A* Event Horizon Telescope Results. VI. Testing the Black Hole Metric}},\ }\href {https://doi.org/10.3847/2041-8213/ac6756} {\bibfield  {journal} {\bibinfo  {journal} {\apjl}\ }\textbf {\bibinfo {volume} {930}},\ \bibinfo {eid} {L17} (\bibinfo {year} {2022}{\natexlab{b}})}\BibitemShut {NoStop}%
\bibitem [{\citenamefont {{Pani}}\ and\ \citenamefont {{Cardoso}}(2009)}]{2009PhRvD..79h4031P}%
  \BibitemOpen
  \bibfield  {author} {\bibinfo {author} {\bibfnamefont {P.}~\bibnamefont {{Pani}}}\ and\ \bibinfo {author} {\bibfnamefont {V.}~\bibnamefont {{Cardoso}}},\ }\bibfield  {title} {\bibinfo {title} {{Are black holes in alternative theories serious astrophysical candidates? The case for Einstein-dilaton-Gauss-Bonnet black holes}},\ }\href {https://doi.org/10.1103/PhysRevD.79.084031} {\bibfield  {journal} {\bibinfo  {journal} {\prd}\ }\textbf {\bibinfo {volume} {79}},\ \bibinfo {eid} {084031} (\bibinfo {year} {2009})},\ \Eprint {https://arxiv.org/abs/0902.1569} {arXiv:0902.1569 [gr-qc]} \BibitemShut {NoStop}%
\bibitem [{\citenamefont {{Shankaranarayanan}}\ and\ \citenamefont {{Johnson}}(2022)}]{2022GReGr..54...44S}%
  \BibitemOpen
  \bibfield  {author} {\bibinfo {author} {\bibfnamefont {S.}~\bibnamefont {{Shankaranarayanan}}}\ and\ \bibinfo {author} {\bibfnamefont {J.~P.}\ \bibnamefont {{Johnson}}},\ }\bibfield  {title} {\bibinfo {title} {{Modified theories of gravity: Why, how and what?}},\ }\href {https://doi.org/10.1007/s10714-022-02927-2} {\bibfield  {journal} {\bibinfo  {journal} {Gen. Relativ. Gravit.}\ }\textbf {\bibinfo {volume} {54}},\ \bibinfo {eid} {44} (\bibinfo {year} {2022})},\ \Eprint {https://arxiv.org/abs/2204.06533} {arXiv:2204.06533 [gr-qc]} \BibitemShut {NoStop}%
\bibitem [{\citenamefont {{Antoniou}}\ \emph {et~al.}(2023)\citenamefont {{Antoniou}}, \citenamefont {{Papageorgiou}},\ and\ \citenamefont {{Kanti}}}]{2023Univ....9..147A}%
  \BibitemOpen
  \bibfield  {author} {\bibinfo {author} {\bibfnamefont {G.}~\bibnamefont {{Antoniou}}}, \bibinfo {author} {\bibfnamefont {A.}~\bibnamefont {{Papageorgiou}}},\ and\ \bibinfo {author} {\bibfnamefont {P.}~\bibnamefont {{Kanti}}},\ }\bibfield  {title} {\bibinfo {title} {{Probing Modified Gravity Theories with Scalar Fields Using Black-Hole Images}},\ }\href {https://doi.org/10.3390/universe9030147} {\bibfield  {journal} {\bibinfo  {journal} {Universe}\ }\textbf {\bibinfo {volume} {9}},\ \bibinfo {eid} {147} (\bibinfo {year} {2023})},\ \Eprint {https://arxiv.org/abs/2210.17533} {arXiv:2210.17533 [gr-qc]} \BibitemShut {NoStop}%
\bibitem [{\citenamefont {{Vagnozzi}}\ \emph {et~al.}(2023)\citenamefont {{Vagnozzi}}, \citenamefont {{Roy}}, \citenamefont {{Tsai}}, \citenamefont {{Visinelli}}, \citenamefont {{Afrin}}, \citenamefont {{Allahyari}}, \citenamefont {{Bambhaniya}}, \citenamefont {{Dey}}, \citenamefont {{Ghosh}}, \citenamefont {{Joshi}}, \citenamefont {{Jusufi}}, \citenamefont {{Khodadi}}, \citenamefont {{Walia}}, \citenamefont {{{\"O}vg{\"u}n}},\ and\ \citenamefont {{Bambi}}}]{2023CQGra..40p5007V}%
  \BibitemOpen
  \bibfield  {author} {\bibinfo {author} {\bibfnamefont {S.}~\bibnamefont {{Vagnozzi}}}, \bibinfo {author} {\bibfnamefont {R.}~\bibnamefont {{Roy}}}, \bibinfo {author} {\bibfnamefont {Y.-D.}\ \bibnamefont {{Tsai}}}, \bibinfo {author} {\bibfnamefont {L.}~\bibnamefont {{Visinelli}}}, \bibinfo {author} {\bibfnamefont {M.}~\bibnamefont {{Afrin}}}, \bibinfo {author} {\bibfnamefont {A.}~\bibnamefont {{Allahyari}}}, \bibinfo {author} {\bibfnamefont {P.}~\bibnamefont {{Bambhaniya}}}, \bibinfo {author} {\bibfnamefont {D.}~\bibnamefont {{Dey}}}, \bibinfo {author} {\bibfnamefont {S.~G.}\ \bibnamefont {{Ghosh}}}, \bibinfo {author} {\bibfnamefont {P.~S.}\ \bibnamefont {{Joshi}}}, \bibinfo {author} {\bibfnamefont {K.}~\bibnamefont {{Jusufi}}}, \bibinfo {author} {\bibfnamefont {M.}~\bibnamefont {{Khodadi}}}, \bibinfo {author} {\bibfnamefont {R.~K.}\ \bibnamefont {{Walia}}}, \bibinfo {author} {\bibfnamefont {A.}~\bibnamefont {{{\"O}vg{\"u}n}}},\ and\ \bibinfo {author} {\bibfnamefont {C.}~\bibnamefont {{Bambi}}},\ }\bibfield
  {title} {\bibinfo {title} {{Horizon-scale tests of gravity theories and fundamental physics from the Event Horizon Telescope image of Sagittarius A (*)}},\ }\href {https://doi.org/10.1088/1361-6382/acd97b} {\bibfield  {journal} {\bibinfo  {journal} {Class. Quantum Gravity}\ }\textbf {\bibinfo {volume} {40}},\ \bibinfo {eid} {165007} (\bibinfo {year} {2023})},\ \Eprint {https://arxiv.org/abs/2205.07787} {arXiv:2205.07787 [gr-qc]} \BibitemShut {NoStop}%
\bibitem [{\citenamefont {{Abishev}}\ \emph {et~al.}(2015)\citenamefont {{Abishev}}, \citenamefont {{Boshkayev}}, \citenamefont {{Dzhunushaliev}},\ and\ \citenamefont {{Ivashchuk}}}]{2015CQGra..32p5010A}%
  \BibitemOpen
  \bibfield  {author} {\bibinfo {author} {\bibfnamefont {M.~E.}\ \bibnamefont {{Abishev}}}, \bibinfo {author} {\bibfnamefont {K.~A.}\ \bibnamefont {{Boshkayev}}}, \bibinfo {author} {\bibfnamefont {V.~D.}\ \bibnamefont {{Dzhunushaliev}}},\ and\ \bibinfo {author} {\bibfnamefont {V.~D.}\ \bibnamefont {{Ivashchuk}}},\ }\bibfield  {title} {\bibinfo {title} {{Dilatonic dyon black hole solutions}},\ }\href {https://doi.org/10.1088/0264-9381/32/16/165010} {\bibfield  {journal} {\bibinfo  {journal} {Class. Quantum Gravity}\ }\textbf {\bibinfo {volume} {32}},\ \bibinfo {eid} {165010} (\bibinfo {year} {2015})},\ \Eprint {https://arxiv.org/abs/1504.07657} {arXiv:1504.07657 [gr-qc]} \BibitemShut {NoStop}%
\bibitem [{\citenamefont {{Abishev}}\ \emph {et~al.}(2017)\citenamefont {{Abishev}}, \citenamefont {{Boshkayev}},\ and\ \citenamefont {{Ivashchuk}}}]{ABI}%
  \BibitemOpen
  \bibfield  {author} {\bibinfo {author} {\bibfnamefont {M.~E.}\ \bibnamefont {{Abishev}}}, \bibinfo {author} {\bibfnamefont {K.~A.}\ \bibnamefont {{Boshkayev}}},\ and\ \bibinfo {author} {\bibfnamefont {V.~D.}\ \bibnamefont {{Ivashchuk}}},\ }\bibfield  {title} {\bibinfo {title} {{Dilatonic dyon-like black hole solutions in the model with two Abelian gauge fields}},\ }\href {https://doi.org/10.1140/epjc/s10052-017-4749-1} {\bibfield  {journal} {\bibinfo  {journal} {Eur. Phys. J. C}\ }\textbf {\bibinfo {volume} {77}},\ \bibinfo {eid} {180} (\bibinfo {year} {2017})},\ \Eprint {https://arxiv.org/abs/1701.02029} {arXiv:1701.02029 [gr-qc]} \BibitemShut {NoStop}%
\bibitem [{\citenamefont {{Belissarova}}\ \emph {et~al.}(2020)\citenamefont {{Belissarova}}, \citenamefont {{Boshkayev}}, \citenamefont {{Ivashchuk}},\ and\ \citenamefont {{Malybayev}}}]{2020JPhCS1690a2143B}%
  \BibitemOpen
  \bibfield  {author} {\bibinfo {author} {\bibfnamefont {F.~B.}\ \bibnamefont {{Belissarova}}}, \bibinfo {author} {\bibfnamefont {K.~A.}\ \bibnamefont {{Boshkayev}}}, \bibinfo {author} {\bibfnamefont {V.~D.}\ \bibnamefont {{Ivashchuk}}},\ and\ \bibinfo {author} {\bibfnamefont {A.~N.}\ \bibnamefont {{Malybayev}}},\ }\bibfield  {title} {\bibinfo {title} {{Special dyon-like black hole solution in the model with two Abelian gauge fields and two scalar fields}},\ }in\ \href {https://doi.org/10.1088/1742-6596/1690/1/012143} {\emph {\bibinfo {booktitle} {J. Phys. Conf. Ser.}}},\ Vol.\ \bibinfo {volume} {1690}\ (\bibinfo {year} {2020})\ p.\ \bibinfo {pages} {012143}\BibitemShut {NoStop}%
\bibitem [{\citenamefont {{Malybayev}}\ \emph {et~al.}(2021)\citenamefont {{Malybayev}}, \citenamefont {{Boshkayev}},\ and\ \citenamefont {{Ivashchuk}}}]{MBI}%
  \BibitemOpen
  \bibfield  {author} {\bibinfo {author} {\bibfnamefont {A.~N.}\ \bibnamefont {{Malybayev}}}, \bibinfo {author} {\bibfnamefont {K.~A.}\ \bibnamefont {{Boshkayev}}},\ and\ \bibinfo {author} {\bibfnamefont {V.~D.}\ \bibnamefont {{Ivashchuk}}},\ }\bibfield  {title} {\bibinfo {title} {{Quasinormal modes in the field of a dyon-like dilatonic black hole}},\ }\href {https://doi.org/10.1140/epjc/s10052-021-09252-z} {\bibfield  {journal} {\bibinfo  {journal} {Eur. Phys. J. C}\ }\textbf {\bibinfo {volume} {81}},\ \bibinfo {eid} {475} (\bibinfo {year} {2021})},\ \Eprint {https://arxiv.org/abs/2103.10920} {arXiv:2103.10920 [gr-qc]} \BibitemShut {NoStop}%
\bibitem [{\citenamefont {{Konoplya}}\ \emph {et~al.}(2019)\citenamefont {{Konoplya}}, \citenamefont {{Zinhailo}},\ and\ \citenamefont {{Stuchl{\'\i}k}}}]{2019PhRvD..99l4042K}%
  \BibitemOpen
  \bibfield  {author} {\bibinfo {author} {\bibfnamefont {R.~A.}\ \bibnamefont {{Konoplya}}}, \bibinfo {author} {\bibfnamefont {A.~F.}\ \bibnamefont {{Zinhailo}}},\ and\ \bibinfo {author} {\bibfnamefont {Z.}~\bibnamefont {{Stuchl{\'\i}k}}},\ }\bibfield  {title} {\bibinfo {title} {{Quasinormal modes, scattering, and Hawking radiation in the vicinity of an Einstein-dilaton-Gauss-Bonnet black hole}},\ }\href {https://doi.org/10.1103/PhysRevD.99.124042} {\bibfield  {journal} {\bibinfo  {journal} {\prd}\ }\textbf {\bibinfo {volume} {99}},\ \bibinfo {eid} {124042} (\bibinfo {year} {2019})},\ \Eprint {https://arxiv.org/abs/1903.03483} {arXiv:1903.03483 [gr-qc]} \BibitemShut {NoStop}%
\bibitem [{\citenamefont {{Uniyal}}\ \emph {et~al.}(2023)\citenamefont {{Uniyal}}, \citenamefont {{Pantig}},\ and\ \citenamefont {{{\"O}vg{\"u}n}}}]{2023PDU....4001178U}%
  \BibitemOpen
  \bibfield  {author} {\bibinfo {author} {\bibfnamefont {A.}~\bibnamefont {{Uniyal}}}, \bibinfo {author} {\bibfnamefont {R.~C.}\ \bibnamefont {{Pantig}}},\ and\ \bibinfo {author} {\bibfnamefont {A.}~\bibnamefont {{{\"O}vg{\"u}n}}},\ }\bibfield  {title} {\bibinfo {title} {{Probing a non-linear electrodynamics black hole with thin accretion disk, shadow, and deflection angle with M87* and Sgr A* from EHT}},\ }\href {https://doi.org/10.1016/j.dark.2023.101178} {\bibfield  {journal} {\bibinfo  {journal} {Phys. Dark Universe}\ }\textbf {\bibinfo {volume} {40}},\ \bibinfo {eid} {101178} (\bibinfo {year} {2023})},\ \Eprint {https://arxiv.org/abs/2205.11072} {arXiv:2205.11072 [gr-qc]} \BibitemShut {NoStop}%
\bibitem [{\citenamefont {{Lambiase}}\ \emph {et~al.}(2023)\citenamefont {{Lambiase}}, \citenamefont {{Pantig}}, \citenamefont {{Jyoti Gogoi}},\ and\ \citenamefont {{{\"O}vg{\"u}n}}}]{2023arXiv230400183L}%
  \BibitemOpen
  \bibfield  {author} {\bibinfo {author} {\bibfnamefont {G.}~\bibnamefont {{Lambiase}}}, \bibinfo {author} {\bibfnamefont {R.~C.}\ \bibnamefont {{Pantig}}}, \bibinfo {author} {\bibfnamefont {D.}~\bibnamefont {{Jyoti Gogoi}}},\ and\ \bibinfo {author} {\bibfnamefont {A.}~\bibnamefont {{{\"O}vg{\"u}n}}},\ }\bibfield  {title} {\bibinfo {title} {{Investigating the Connection between Generalized Uncertainty Principle and Asymptotically Safe Gravity in Black Hole Signatures through Shadow and Quasinormal Modes}},\ }\href {https://doi.org/10.48550/arXiv.2304.00183} {\bibfield  {journal} {\bibinfo  {journal} {arXiv e-prints}\ ,\ \bibinfo {eid} {arXiv:2304.00183}} (\bibinfo {year} {2023})},\ \Eprint {https://arxiv.org/abs/2304.00183} {arXiv:2304.00183 [gr-qc]} \BibitemShut {NoStop}%
\bibitem [{\citenamefont {{Boshkayev}}\ \emph {et~al.}(2021)\citenamefont {{Boshkayev}}, \citenamefont {{Konysbayev}}, \citenamefont {{Kurmanov}}, \citenamefont {{Luongo}}, \citenamefont {{Malafarina}},\ and\ \citenamefont {{Quevedo}}}]{2021PhRvD.104h4009B}%
  \BibitemOpen
  \bibfield  {author} {\bibinfo {author} {\bibfnamefont {K.}~\bibnamefont {{Boshkayev}}}, \bibinfo {author} {\bibfnamefont {T.}~\bibnamefont {{Konysbayev}}}, \bibinfo {author} {\bibfnamefont {E.}~\bibnamefont {{Kurmanov}}}, \bibinfo {author} {\bibfnamefont {O.}~\bibnamefont {{Luongo}}}, \bibinfo {author} {\bibfnamefont {D.}~\bibnamefont {{Malafarina}}},\ and\ \bibinfo {author} {\bibfnamefont {H.}~\bibnamefont {{Quevedo}}},\ }\bibfield  {title} {\bibinfo {title} {{Luminosity of accretion disks in compact objects with a quadrupole}},\ }\href {https://doi.org/10.1103/PhysRevD.104.084009} {\bibfield  {journal} {\bibinfo  {journal} {\prd}\ }\textbf {\bibinfo {volume} {104}},\ \bibinfo {eid} {084009} (\bibinfo {year} {2021})},\ \Eprint {https://arxiv.org/abs/2106.04932} {arXiv:2106.04932 [gr-qc]} \BibitemShut {NoStop}%
\bibitem [{\citenamefont {{Boshkayev}}\ \emph {et~al.}(2016{\natexlab{a}})\citenamefont {{Boshkayev}}, \citenamefont {{Gasper{\'\i}n}}, \citenamefont {{Guti{\'e}rrez-Pi{\~n}eres}}, \citenamefont {{Quevedo}},\ and\ \citenamefont {{Toktarbay}}}]{2016PhRvD..93b4024B}%
  \BibitemOpen
  \bibfield  {author} {\bibinfo {author} {\bibfnamefont {K.}~\bibnamefont {{Boshkayev}}}, \bibinfo {author} {\bibfnamefont {E.}~\bibnamefont {{Gasper{\'\i}n}}}, \bibinfo {author} {\bibfnamefont {A.~C.}\ \bibnamefont {{Guti{\'e}rrez-Pi{\~n}eres}}}, \bibinfo {author} {\bibfnamefont {H.}~\bibnamefont {{Quevedo}}},\ and\ \bibinfo {author} {\bibfnamefont {S.}~\bibnamefont {{Toktarbay}}},\ }\bibfield  {title} {\bibinfo {title} {{Motion of test particles in the field of a naked singularity}},\ }\href {https://doi.org/10.1103/PhysRevD.93.024024} {\bibfield  {journal} {\bibinfo  {journal} {\prd}\ }\textbf {\bibinfo {volume} {93}},\ \bibinfo {eid} {024024} (\bibinfo {year} {2016}{\natexlab{a}})},\ \Eprint {https://arxiv.org/abs/1509.03827} {arXiv:1509.03827 [gr-qc]} \BibitemShut {NoStop}%
\bibitem [{\citenamefont {{Bini}}\ \emph {et~al.}(2013)\citenamefont {{Bini}}, \citenamefont {{Boshkayev}}, \citenamefont {{Ruffini}},\ and\ \citenamefont {{Siutsou}}}]{2013NCimC..36S..31B}%
  \BibitemOpen
  \bibfield  {author} {\bibinfo {author} {\bibfnamefont {D.}~\bibnamefont {{Bini}}}, \bibinfo {author} {\bibfnamefont {K.}~\bibnamefont {{Boshkayev}}}, \bibinfo {author} {\bibfnamefont {R.}~\bibnamefont {{Ruffini}}},\ and\ \bibinfo {author} {\bibfnamefont {I.}~\bibnamefont {{Siutsou}}},\ }\bibfield  {title} {\bibinfo {title} {{Equatorial circular geodesics in the Hartle-Thorne spacetime}},\ }\href {https://doi.org/10.1393/ncc/i2013-11483-8} {\bibfield  {journal} {\bibinfo  {journal} {Nuovo Cimento C Geophysics Space Physics C}\ }\textbf {\bibinfo {volume} {36}},\ \bibinfo {pages} {31} (\bibinfo {year} {2013})},\ \Eprint {https://arxiv.org/abs/1306.4792} {arXiv:1306.4792 [gr-qc]} \BibitemShut {NoStop}%
\bibitem [{\citenamefont {{Boshkayev}}\ \emph {et~al.}(2016{\natexlab{b}})\citenamefont {{Boshkayev}}, \citenamefont {{Quevedo}}, \citenamefont {{Abutalip}}, \citenamefont {{Kalymova}},\ and\ \citenamefont {{Suleymanova}}}]{2016IJMPA..3141006B}%
  \BibitemOpen
  \bibfield  {author} {\bibinfo {author} {\bibfnamefont {K.~A.}\ \bibnamefont {{Boshkayev}}}, \bibinfo {author} {\bibfnamefont {H.}~\bibnamefont {{Quevedo}}}, \bibinfo {author} {\bibfnamefont {M.~S.}\ \bibnamefont {{Abutalip}}}, \bibinfo {author} {\bibfnamefont {Z.~A.}\ \bibnamefont {{Kalymova}}},\ and\ \bibinfo {author} {\bibfnamefont {S.~S.}\ \bibnamefont {{Suleymanova}}},\ }\bibfield  {title} {\bibinfo {title} {{Geodesics in the field of a rotating deformed gravitational source}},\ }\href {https://doi.org/10.1142/S0217751X16410062} {\bibfield  {journal} {\bibinfo  {journal} {International Journal of Modern Physics A}\ }\textbf {\bibinfo {volume} {31}},\ \bibinfo {eid} {1641006} (\bibinfo {year} {2016}{\natexlab{b}})},\ \Eprint {https://arxiv.org/abs/1510.02016} {arXiv:1510.02016 [gr-qc]} \BibitemShut {NoStop}%
\bibitem [{\citenamefont {{Panah}}\ \emph {et~al.}(2018)\citenamefont {{Panah}}, \citenamefont {{Hendi}}, \citenamefont {{Panahiyan}},\ and\ \citenamefont {{Hassaine}}}]{2018PhRvD..98h4006P}%
  \BibitemOpen
  \bibfield  {author} {\bibinfo {author} {\bibfnamefont {B.~E.}\ \bibnamefont {{Panah}}}, \bibinfo {author} {\bibfnamefont {S.~H.}\ \bibnamefont {{Hendi}}}, \bibinfo {author} {\bibfnamefont {S.}~\bibnamefont {{Panahiyan}}},\ and\ \bibinfo {author} {\bibfnamefont {M.}~\bibnamefont {{Hassaine}}},\ }\bibfield  {title} {\bibinfo {title} {{BTZ dilatonic black holes coupled to Maxwell and Born-Infeld electrodynamics}},\ }\href {https://doi.org/10.1103/PhysRevD.98.084006} {\bibfield  {journal} {\bibinfo  {journal} {\prd}\ }\textbf {\bibinfo {volume} {98}},\ \bibinfo {eid} {084006} (\bibinfo {year} {2018})}\BibitemShut {NoStop}%
\bibitem [{\citenamefont {{Hendi}}\ \emph {et~al.}(2017{\natexlab{a}})\citenamefont {{Hendi}}, \citenamefont {{Panah}}, \citenamefont {{Panahiyan}},\ and\ \citenamefont {{Momennia}}}]{2017EPJC...77..647H}%
  \BibitemOpen
  \bibfield  {author} {\bibinfo {author} {\bibfnamefont {S.~H.}\ \bibnamefont {{Hendi}}}, \bibinfo {author} {\bibfnamefont {B.~E.}\ \bibnamefont {{Panah}}}, \bibinfo {author} {\bibfnamefont {S.}~\bibnamefont {{Panahiyan}}},\ and\ \bibinfo {author} {\bibfnamefont {M.}~\bibnamefont {{Momennia}}},\ }\bibfield  {title} {\bibinfo {title} {{Dilatonic black holes in gravity's rainbow with a nonlinear source: the effects of thermal fluctuations}},\ }\href {https://doi.org/10.1140/epjc/s10052-017-5211-0} {\bibfield  {journal} {\bibinfo  {journal} {European Physical Journal C}\ }\textbf {\bibinfo {volume} {77}},\ \bibinfo {eid} {647} (\bibinfo {year} {2017}{\natexlab{a}})},\ \Eprint {https://arxiv.org/abs/1708.06634} {arXiv:1708.06634 [gr-qc]} \BibitemShut {NoStop}%
\bibitem [{\citenamefont {{Hendi}}\ \emph {et~al.}(2017{\natexlab{b}})\citenamefont {{Hendi}}, \citenamefont {{Eslam Panah}}, \citenamefont {{Panahiyan}},\ and\ \citenamefont {{Sheykhi}}}]{2017PhLB..767..214H}%
  \BibitemOpen
  \bibfield  {author} {\bibinfo {author} {\bibfnamefont {S.~H.}\ \bibnamefont {{Hendi}}}, \bibinfo {author} {\bibfnamefont {B.}~\bibnamefont {{Eslam Panah}}}, \bibinfo {author} {\bibfnamefont {S.}~\bibnamefont {{Panahiyan}}},\ and\ \bibinfo {author} {\bibfnamefont {A.}~\bibnamefont {{Sheykhi}}},\ }\bibfield  {title} {\bibinfo {title} {{Dilatonic BTZ black holes with power-law field}},\ }\href {https://doi.org/10.1016/j.physletb.2017.01.066} {\bibfield  {journal} {\bibinfo  {journal} {Physics Letters B}\ }\textbf {\bibinfo {volume} {767}},\ \bibinfo {pages} {214} (\bibinfo {year} {2017}{\natexlab{b}})},\ \Eprint {https://arxiv.org/abs/1703.03403} {arXiv:1703.03403 [gr-qc]} \BibitemShut {NoStop}%
\bibitem [{\citenamefont {{Hendi}}\ \emph {et~al.}(2016)\citenamefont {{Hendi}}, \citenamefont {{Faizal}}, \citenamefont {{Panah}},\ and\ \citenamefont {{Panahiyan}}}]{2016EPJC...76..296H}%
  \BibitemOpen
  \bibfield  {author} {\bibinfo {author} {\bibfnamefont {S.~H.}\ \bibnamefont {{Hendi}}}, \bibinfo {author} {\bibfnamefont {M.}~\bibnamefont {{Faizal}}}, \bibinfo {author} {\bibfnamefont {B.~E.}\ \bibnamefont {{Panah}}},\ and\ \bibinfo {author} {\bibfnamefont {S.}~\bibnamefont {{Panahiyan}}},\ }\bibfield  {title} {\bibinfo {title} {{Charged dilatonic black holes in gravity's rainbow}},\ }\href {https://doi.org/10.1140/epjc/s10052-016-4119-4} {\bibfield  {journal} {\bibinfo  {journal} {European Physical Journal C}\ }\textbf {\bibinfo {volume} {76}},\ \bibinfo {eid} {296} (\bibinfo {year} {2016})},\ \Eprint {https://arxiv.org/abs/1508.00234} {arXiv:1508.00234 [hep-th]} \BibitemShut {NoStop}%
\bibitem [{\citenamefont {{Hendi}}\ \emph {et~al.}(2015)\citenamefont {{Hendi}}, \citenamefont {{Sheykhi}}, \citenamefont {{Panahiyan}},\ and\ \citenamefont {{Eslam Panah}}}]{2015PhRvD..92f4028H}%
  \BibitemOpen
  \bibfield  {author} {\bibinfo {author} {\bibfnamefont {S.~H.}\ \bibnamefont {{Hendi}}}, \bibinfo {author} {\bibfnamefont {A.}~\bibnamefont {{Sheykhi}}}, \bibinfo {author} {\bibfnamefont {S.}~\bibnamefont {{Panahiyan}}},\ and\ \bibinfo {author} {\bibfnamefont {B.}~\bibnamefont {{Eslam Panah}}},\ }\bibfield  {title} {\bibinfo {title} {{Phase transition and thermodynamic geometry of Einstein-Maxwell-dilaton black holes}},\ }\href {https://doi.org/10.1103/PhysRevD.92.064028} {\bibfield  {journal} {\bibinfo  {journal} {\prd}\ }\textbf {\bibinfo {volume} {92}},\ \bibinfo {eid} {064028} (\bibinfo {year} {2015})},\ \Eprint {https://arxiv.org/abs/1509.08593} {arXiv:1509.08593 [hep-th]} \BibitemShut {NoStop}%
\bibitem [{\citenamefont {{Cveti{\v{c}}}}\ \emph {et~al.}(2016)\citenamefont {{Cveti{\v{c}}}}, \citenamefont {{Gibbons}},\ and\ \citenamefont {{Pope}}}]{CGP}%
  \BibitemOpen
  \bibfield  {author} {\bibinfo {author} {\bibfnamefont {M.}~\bibnamefont {{Cveti{\v{c}}}}}, \bibinfo {author} {\bibfnamefont {G.~W.}\ \bibnamefont {{Gibbons}}},\ and\ \bibinfo {author} {\bibfnamefont {C.~N.}\ \bibnamefont {{Pope}}},\ }\bibfield  {title} {\bibinfo {title} {{Photon spheres and sonic horizons in black holes from supergravity and other theories}},\ }\href {https://doi.org/10.1103/PhysRevD.94.106005} {\bibfield  {journal} {\bibinfo  {journal} {\prd}\ }\textbf {\bibinfo {volume} {94}},\ \bibinfo {eid} {106005} (\bibinfo {year} {2016})},\ \Eprint {https://arxiv.org/abs/1608.02202} {arXiv:1608.02202 [gr-qc]} \BibitemShut {NoStop}%
\bibitem [{\citenamefont {{Huang}}\ and\ \citenamefont {{Zhang}}(2022)}]{2022PhRvD.105l4056H}%
  \BibitemOpen
  \bibfield  {author} {\bibinfo {author} {\bibfnamefont {Y.}~\bibnamefont {{Huang}}}\ and\ \bibinfo {author} {\bibfnamefont {H.}~\bibnamefont {{Zhang}}},\ }\bibfield  {title} {\bibinfo {title} {{True gravitational atoms: Spherical cloud of dilatonic black holes}},\ }\href {https://doi.org/10.1103/PhysRevD.105.124056} {\bibfield  {journal} {\bibinfo  {journal} {\prd}\ }\textbf {\bibinfo {volume} {105}},\ \bibinfo {eid} {124056} (\bibinfo {year} {2022})},\ \Eprint {https://arxiv.org/abs/2206.05645} {arXiv:2206.05645 [gr-qc]} \BibitemShut {NoStop}%
\bibitem [{\citenamefont {{Huang}}\ and\ \citenamefont {{Zhang}}(2020)}]{2020EPJC...80..654H}%
  \BibitemOpen
  \bibfield  {author} {\bibinfo {author} {\bibfnamefont {Y.}~\bibnamefont {{Huang}}}\ and\ \bibinfo {author} {\bibfnamefont {H.}~\bibnamefont {{Zhang}}},\ }\bibfield  {title} {\bibinfo {title} {{Scattering of massless scalar field by charged dilatonic black holes}},\ }\href {https://doi.org/10.1140/epjc/s10052-020-8228-8} {\bibfield  {journal} {\bibinfo  {journal} {European Physical Journal C}\ }\textbf {\bibinfo {volume} {80}},\ \bibinfo {eid} {654} (\bibinfo {year} {2020})},\ \Eprint {https://arxiv.org/abs/2006.01388} {arXiv:2006.01388 [gr-qc]} \BibitemShut {NoStop}%
\bibitem [{\citenamefont {{Cadoni}}\ \emph {et~al.}(2010)\citenamefont {{Cadoni}}, \citenamefont {{D'Appollonio}},\ and\ \citenamefont {{Pani}}}]{2010JHEP...03..100C}%
  \BibitemOpen
  \bibfield  {author} {\bibinfo {author} {\bibfnamefont {M.}~\bibnamefont {{Cadoni}}}, \bibinfo {author} {\bibfnamefont {G.}~\bibnamefont {{D'Appollonio}}},\ and\ \bibinfo {author} {\bibfnamefont {P.}~\bibnamefont {{Pani}}},\ }\bibfield  {title} {\bibinfo {title} {{Phase transitions between Reissner-Nordstrom and dilatonic black holes in 4D AdS spacetime}},\ }\href {https://doi.org/10.1007/JHEP03(2010)100} {\bibfield  {journal} {\bibinfo  {journal} {Journal of High Energy Physics}\ }\textbf {\bibinfo {volume} {2010}},\ \bibinfo {eid} {100} (\bibinfo {year} {2010})},\ \Eprint {https://arxiv.org/abs/0912.3520} {arXiv:0912.3520 [hep-th]} \BibitemShut {NoStop}%
\bibitem [{\citenamefont {{Chen}}(2008)}]{2008PThPS.172..161C}%
  \BibitemOpen
  \bibfield  {author} {\bibinfo {author} {\bibfnamefont {C.}~\bibnamefont {{Chen}}},\ }\bibfield  {title} {\bibinfo {title} {{Extremal Dilatonic Black Holes in 4D Gauss-Bonnet Gravity}},\ }\href {https://doi.org/10.1143/PTPS.172.161} {\bibfield  {journal} {\bibinfo  {journal} {Progress of Theoretical Physics Supplement}\ }\textbf {\bibinfo {volume} {172}},\ \bibinfo {pages} {161} (\bibinfo {year} {2008})},\ \Eprint {https://arxiv.org/abs/0801.0032} {arXiv:0801.0032 [hep-th]} \BibitemShut {NoStop}%
\bibitem [{\citenamefont {{Li}}\ \emph {et~al.}(2022)\citenamefont {{Li}}, \citenamefont {{Mirzaev}}, \citenamefont {{Abdujabbarov}}, \citenamefont {{Malafarina}}, \citenamefont {{Ahmedov}},\ and\ \citenamefont {{Han}}}]{2022PhRvD.106h4041L}%
  \BibitemOpen
  \bibfield  {author} {\bibinfo {author} {\bibfnamefont {S.}~\bibnamefont {{Li}}}, \bibinfo {author} {\bibfnamefont {T.}~\bibnamefont {{Mirzaev}}}, \bibinfo {author} {\bibfnamefont {A.~A.}\ \bibnamefont {{Abdujabbarov}}}, \bibinfo {author} {\bibfnamefont {D.}~\bibnamefont {{Malafarina}}}, \bibinfo {author} {\bibfnamefont {B.}~\bibnamefont {{Ahmedov}}},\ and\ \bibinfo {author} {\bibfnamefont {W.-B.}\ \bibnamefont {{Han}}},\ }\bibfield  {title} {\bibinfo {title} {{Constraining the deformation of a rotating black hole mimicker from its shadow}},\ }\href {https://doi.org/10.1103/PhysRevD.106.084041} {\bibfield  {journal} {\bibinfo  {journal} {\prd}\ }\textbf {\bibinfo {volume} {106}},\ \bibinfo {eid} {084041} (\bibinfo {year} {2022})},\ \Eprint {https://arxiv.org/abs/2207.10933} {arXiv:2207.10933 [gr-qc]} \BibitemShut {NoStop}%
\bibitem [{\citenamefont {{Sarikulov}}\ \emph {et~al.}(2022)\citenamefont {{Sarikulov}}, \citenamefont {{Atamurotov}}, \citenamefont {{Abdujabbarov}},\ and\ \citenamefont {{Ahmedov}}}]{2022EPJC...82..771S}%
  \BibitemOpen
  \bibfield  {author} {\bibinfo {author} {\bibfnamefont {F.}~\bibnamefont {{Sarikulov}}}, \bibinfo {author} {\bibfnamefont {F.}~\bibnamefont {{Atamurotov}}}, \bibinfo {author} {\bibfnamefont {A.}~\bibnamefont {{Abdujabbarov}}},\ and\ \bibinfo {author} {\bibfnamefont {B.}~\bibnamefont {{Ahmedov}}},\ }\bibfield  {title} {\bibinfo {title} {{Shadow of the Kerr-like black hole}},\ }\href {https://doi.org/10.1140/epjc/s10052-022-10711-4} {\bibfield  {journal} {\bibinfo  {journal} {European Physical Journal C}\ }\textbf {\bibinfo {volume} {82}},\ \bibinfo {eid} {771} (\bibinfo {year} {2022})}\BibitemShut {NoStop}%
\bibitem [{\citenamefont {{Turimov}}\ \emph {et~al.}(2022)\citenamefont {{Turimov}}, \citenamefont {{Boboqambarova}}, \citenamefont {{Ahmedov}},\ and\ \citenamefont {{Stuchl{\'\i}k}}}]{2022EPJP..137..222T}%
  \BibitemOpen
  \bibfield  {author} {\bibinfo {author} {\bibfnamefont {B.}~\bibnamefont {{Turimov}}}, \bibinfo {author} {\bibfnamefont {M.}~\bibnamefont {{Boboqambarova}}}, \bibinfo {author} {\bibfnamefont {B.}~\bibnamefont {{Ahmedov}}},\ and\ \bibinfo {author} {\bibfnamefont {Z.}~\bibnamefont {{Stuchl{\'\i}k}}},\ }\bibfield  {title} {\bibinfo {title} {{Distinguishable feature of electric and magnetic charged black hole}},\ }\href {https://doi.org/10.1140/epjp/s13360-022-02390-7} {\bibfield  {journal} {\bibinfo  {journal} {European Physical Journal Plus}\ }\textbf {\bibinfo {volume} {137}},\ \bibinfo {eid} {222} (\bibinfo {year} {2022})}\BibitemShut {NoStop}%
\bibitem [{\citenamefont {{Zeng}}\ \emph {et~al.}(2006)\citenamefont {{Zeng}}, \citenamefont {{L{\"u}}},\ and\ \citenamefont {{Wang}}}]{2006ChPhL..23.1648Z}%
  \BibitemOpen
  \bibfield  {author} {\bibinfo {author} {\bibfnamefont {Y.}~\bibnamefont {{Zeng}}}, \bibinfo {author} {\bibfnamefont {J.-L.}\ \bibnamefont {{L{\"u}}}},\ and\ \bibinfo {author} {\bibfnamefont {Y.-J.}\ \bibnamefont {{Wang}}},\ }\bibfield  {title} {\bibinfo {title} {{Geodesics of Spherical Dilaton Spacetimes}},\ }\href {https://doi.org/10.1088/0256-307X/23/6/081} {\bibfield  {journal} {\bibinfo  {journal} {Chinese Physics Letters}\ }\textbf {\bibinfo {volume} {23}},\ \bibinfo {pages} {1648} (\bibinfo {year} {2006})}\BibitemShut {NoStop}%
\bibitem [{\citenamefont {{Blaga}}(2015)}]{2015SerAJ.190...41B}%
  \BibitemOpen
  \bibfield  {author} {\bibinfo {author} {\bibfnamefont {C.}~\bibnamefont {{Blaga}}},\ }\bibfield  {title} {\bibinfo {title} {{Timelike Geodesics Around a Charged Spherically Symmetric Dilaton Black Hole}},\ }\href {https://doi.org/10.2298/SAJ1590041B} {\bibfield  {journal} {\bibinfo  {journal} {Serbian Astronomical Journal}\ }\textbf {\bibinfo {volume} {190}},\ \bibinfo {pages} {41} (\bibinfo {year} {2015})},\ \Eprint {https://arxiv.org/abs/1407.1504} {arXiv:1407.1504 [gr-qc]} \BibitemShut {NoStop}%
\bibitem [{\citenamefont {{Sarkar}}\ \emph {et~al.}(2018)\citenamefont {{Sarkar}}, \citenamefont {{Rahaman}}, \citenamefont {{Radinschi}}, \citenamefont {{Grammenos}},\ and\ \citenamefont {{Chakraborty}}}]{2018arXiv180500295S}%
  \BibitemOpen
  \bibfield  {author} {\bibinfo {author} {\bibfnamefont {S.}~\bibnamefont {{Sarkar}}}, \bibinfo {author} {\bibfnamefont {F.}~\bibnamefont {{Rahaman}}}, \bibinfo {author} {\bibfnamefont {I.}~\bibnamefont {{Radinschi}}}, \bibinfo {author} {\bibfnamefont {T.}~\bibnamefont {{Grammenos}}},\ and\ \bibinfo {author} {\bibfnamefont {J.}~\bibnamefont {{Chakraborty}}},\ }\bibfield  {title} {\bibinfo {title} {{Particle motion around charged black holes in generalized dilaton-axion gravity}},\ }\href {https://doi.org/10.1155/2018/5427158} {\bibfield  {journal} {\bibinfo  {journal} {Advances in High Energy Physics}\ }\textbf {\bibinfo {volume} {2018}},\ \bibinfo {eid} {5427158} (\bibinfo {year} {2018})},\ \Eprint {https://arxiv.org/abs/1805.00295} {arXiv:1805.00295 [gr-qc]} \BibitemShut {NoStop}%
\bibitem [{\citenamefont {{Blaga}}\ \emph {et~al.}(2023)\citenamefont {{Blaga}}, \citenamefont {{Blaga}},\ and\ \citenamefont {{Harko}}}]{2023Symm...15..329B}%
  \BibitemOpen
  \bibfield  {author} {\bibinfo {author} {\bibfnamefont {C.}~\bibnamefont {{Blaga}}}, \bibinfo {author} {\bibfnamefont {P.}~\bibnamefont {{Blaga}}},\ and\ \bibinfo {author} {\bibfnamefont {T.}~\bibnamefont {{Harko}}},\ }\bibfield  {title} {\bibinfo {title} {{Jacobi and Lyapunov Stability Analysis of Circular Geodesics around a Spherically Symmetric Dilaton Black Hole}},\ }\href {https://doi.org/10.3390/sym15020329} {\bibfield  {journal} {\bibinfo  {journal} {Symmetry}\ }\textbf {\bibinfo {volume} {15}},\ \bibinfo {pages} {329} (\bibinfo {year} {2023})},\ \Eprint {https://arxiv.org/abs/2301.07678} {arXiv:2301.07678 [gr-qc]} \BibitemShut {NoStop}%
\bibitem [{\citenamefont {{Heydari-Fard}}\ \emph {et~al.}(2022)\citenamefont {{Heydari-Fard}}, \citenamefont {{Heydari-Fard}},\ and\ \citenamefont {{Sepangi}}}]{2022PhRvD.105l4009H}%
  \BibitemOpen
  \bibfield  {author} {\bibinfo {author} {\bibfnamefont {M.}~\bibnamefont {{Heydari-Fard}}}, \bibinfo {author} {\bibfnamefont {M.}~\bibnamefont {{Heydari-Fard}}},\ and\ \bibinfo {author} {\bibfnamefont {H.~R.}\ \bibnamefont {{Sepangi}}},\ }\bibfield  {title} {\bibinfo {title} {{Null geodesics and shadow of hairy black holes in Einstein-Maxwell-dilaton gravity}},\ }\href {https://doi.org/10.1103/PhysRevD.105.124009} {\bibfield  {journal} {\bibinfo  {journal} {\prd}\ }\textbf {\bibinfo {volume} {105}},\ \bibinfo {eid} {124009} (\bibinfo {year} {2022})},\ \Eprint {https://arxiv.org/abs/2110.02713} {arXiv:2110.02713 [gr-qc]} \BibitemShut {NoStop}%
\bibitem [{\citenamefont {{Soroushfar}}\ \emph {et~al.}(2016)\citenamefont {{Soroushfar}}, \citenamefont {{Saffari}},\ and\ \citenamefont {{Sahami}}}]{2016PhRvD..94b4010S}%
  \BibitemOpen
  \bibfield  {author} {\bibinfo {author} {\bibfnamefont {S.}~\bibnamefont {{Soroushfar}}}, \bibinfo {author} {\bibfnamefont {R.}~\bibnamefont {{Saffari}}},\ and\ \bibinfo {author} {\bibfnamefont {E.}~\bibnamefont {{Sahami}}},\ }\bibfield  {title} {\bibinfo {title} {{Geodesic equations in the static and rotating dilaton black holes: Analytical solutions and applications}},\ }\href {https://doi.org/10.1103/PhysRevD.94.024010} {\bibfield  {journal} {\bibinfo  {journal} {\prd}\ }\textbf {\bibinfo {volume} {94}},\ \bibinfo {eid} {024010} (\bibinfo {year} {2016})},\ \Eprint {https://arxiv.org/abs/1601.03143} {arXiv:1601.03143 [gr-qc]} \BibitemShut {NoStop}%
\bibitem [{\citenamefont {{Flathmann}}\ and\ \citenamefont {{Grunau}}(2015)}]{2015PhRvD..92j4027F}%
  \BibitemOpen
  \bibfield  {author} {\bibinfo {author} {\bibfnamefont {K.}~\bibnamefont {{Flathmann}}}\ and\ \bibinfo {author} {\bibfnamefont {S.}~\bibnamefont {{Grunau}}},\ }\bibfield  {title} {\bibinfo {title} {{Analytic solutions of the geodesic equation for Einstein-Maxwell-dilaton-axion black holes}},\ }\href {https://doi.org/10.1103/PhysRevD.92.104027} {\bibfield  {journal} {\bibinfo  {journal} {\prd}\ }\textbf {\bibinfo {volume} {92}},\ \bibinfo {eid} {104027} (\bibinfo {year} {2015})},\ \Eprint {https://arxiv.org/abs/1509.03135} {arXiv:1509.03135 [gr-qc]} \BibitemShut {NoStop}%
\bibitem [{\citenamefont {{Ivashchuk}}\ \emph {et~al.}(2023)\citenamefont {{Ivashchuk}}, \citenamefont {{Malybayev}}, \citenamefont {{Nurbakova}},\ and\ \citenamefont {{Takey}}}]{IMNT}%
  \BibitemOpen
  \bibfield  {author} {\bibinfo {author} {\bibfnamefont {V.~D.}\ \bibnamefont {{Ivashchuk}}}, \bibinfo {author} {\bibfnamefont {A.~N.}\ \bibnamefont {{Malybayev}}}, \bibinfo {author} {\bibfnamefont {G.~S.}\ \bibnamefont {{Nurbakova}}},\ and\ \bibinfo {author} {\bibfnamefont {G.}~\bibnamefont {{Takey}}},\ }\bibfield  {title} {\bibinfo {title} {{Photon Spheres near Dilatonic Dyon-Like Black Holes in a Model with Two Abelian Gauge Fields and Two Scalar Fields}},\ }\href {https://doi.org/10.1134/S0202289323040114} {\bibfield  {journal} {\bibinfo  {journal} {Gravitation and Cosmology}\ }\textbf {\bibinfo {volume} {29}},\ \bibinfo {pages} {411} (\bibinfo {year} {2023})}\BibitemShut {NoStop}%
\bibitem [{\citenamefont {{Boshkayev}}\ \emph {et~al.}(2024)\citenamefont {{Boshkayev}}, \citenamefont {{Suliyeva}}, \citenamefont {{Ivashchuk}},\ and\ \citenamefont {{Urazalina}}}]{BSIU}%
  \BibitemOpen
  \bibfield  {author} {\bibinfo {author} {\bibfnamefont {K.}~\bibnamefont {{Boshkayev}}}, \bibinfo {author} {\bibfnamefont {G.}~\bibnamefont {{Suliyeva}}}, \bibinfo {author} {\bibfnamefont {V.}~\bibnamefont {{Ivashchuk}}},\ and\ \bibinfo {author} {\bibfnamefont {A.}~\bibnamefont {{Urazalina}}},\ }\bibfield  {title} {\bibinfo {title} {{Circular geodesics in the field of double-charged dilatonic black holes}},\ }\href {https://doi.org/10.1140/epjc/s10052-023-12337-6} {\bibfield  {journal} {\bibinfo  {journal} {European Physical Journal C}\ }\textbf {\bibinfo {volume} {84}},\ \bibinfo {eid} {19} (\bibinfo {year} {2024})},\ \Eprint {https://arxiv.org/abs/2306.01927} {arXiv:2306.01927 [gr-qc]} \BibitemShut {NoStop}%
\bibitem [{\citenamefont {{Stuchl{\'\i}k}}\ \emph {et~al.}(2020)\citenamefont {{Stuchl{\'\i}k}}, \citenamefont {{Kolo{\v{s}}}}, \citenamefont {{Kov{\'a}{\v{r}}}}, \citenamefont {{Slan{\'y}}},\ and\ \citenamefont {{Tursunov}}}]{2020Univ....6...26S}%
  \BibitemOpen
  \bibfield  {author} {\bibinfo {author} {\bibfnamefont {Z.}~\bibnamefont {{Stuchl{\'\i}k}}}, \bibinfo {author} {\bibfnamefont {M.}~\bibnamefont {{Kolo{\v{s}}}}}, \bibinfo {author} {\bibfnamefont {J.}~\bibnamefont {{Kov{\'a}{\v{r}}}}}, \bibinfo {author} {\bibfnamefont {P.}~\bibnamefont {{Slan{\'y}}}},\ and\ \bibinfo {author} {\bibfnamefont {A.}~\bibnamefont {{Tursunov}}},\ }\bibfield  {title} {\bibinfo {title} {{Influence of Cosmic Repulsion and Magnetic Fields on Accretion Disks Rotating around Kerr Black Holes}},\ }\href {https://doi.org/10.3390/universe6020026} {\bibfield  {journal} {\bibinfo  {journal} {Universe}\ }\textbf {\bibinfo {volume} {6}},\ \bibinfo {pages} {26} (\bibinfo {year} {2020})}\BibitemShut {NoStop}%
\bibitem [{\citenamefont {{Bronnikov}}\ and\ \citenamefont {{Shikin}}(1977)}]{BS}%
  \BibitemOpen
  \bibfield  {author} {\bibinfo {author} {\bibfnamefont {K.~A.}\ \bibnamefont {{Bronnikov}}}\ and\ \bibinfo {author} {\bibfnamefont {G.~N.}\ \bibnamefont {{Shikin}}},\ }\bibfield  {title} {\bibinfo {title} {{Interacting fields in general relativity theory}},\ }\href {https://doi.org/10.1007/BF00897114} {\bibfield  {journal} {\bibinfo  {journal} {Soviet Physics Journal}\ }\textbf {\bibinfo {volume} {20}},\ \bibinfo {pages} {1138} (\bibinfo {year} {1977})}\BibitemShut {NoStop}%
\bibitem [{\citenamefont {{Gibbons}}\ and\ \citenamefont {{Maeda}}(1988)}]{1988NuPhB.298..741G}%
  \BibitemOpen
  \bibfield  {author} {\bibinfo {author} {\bibfnamefont {G.~W.}\ \bibnamefont {{Gibbons}}}\ and\ \bibinfo {author} {\bibfnamefont {K.-I.}\ \bibnamefont {{Maeda}}},\ }\bibfield  {title} {\bibinfo {title} {{Black holes and membranes in higher-dimensional theories with dilaton fields}},\ }\href {https://doi.org/10.1016/0550-3213(88)90006-5} {\bibfield  {journal} {\bibinfo  {journal} {Nuclear Physics B}\ }\textbf {\bibinfo {volume} {298}},\ \bibinfo {pages} {741} (\bibinfo {year} {1988})}\BibitemShut {NoStop}%
\bibitem [{\citenamefont {{Garfinkle}}\ \emph {et~al.}(1991)\citenamefont {{Garfinkle}}, \citenamefont {{Horowitz}},\ and\ \citenamefont {{Strominger}}}]{1991PhRvD..43.3140G}%
  \BibitemOpen
  \bibfield  {author} {\bibinfo {author} {\bibfnamefont {D.}~\bibnamefont {{Garfinkle}}}, \bibinfo {author} {\bibfnamefont {G.~T.}\ \bibnamefont {{Horowitz}}},\ and\ \bibinfo {author} {\bibfnamefont {A.}~\bibnamefont {{Strominger}}},\ }\bibfield  {title} {\bibinfo {title} {{Charged black holes in string theory}},\ }\href {https://doi.org/10.1103/PhysRevD.43.3140} {\bibfield  {journal} {\bibinfo  {journal} {\prd}\ }\textbf {\bibinfo {volume} {43}},\ \bibinfo {pages} {3140} (\bibinfo {year} {1991})}\BibitemShut {NoStop}%
\bibitem [{\citenamefont {{Garfinkle}}\ \emph {et~al.}(1992)\citenamefont {{Garfinkle}}, \citenamefont {{Horowitz}},\ and\ \citenamefont {{Strominger}}}]{1992PhRvD..45.3888G}%
  \BibitemOpen
  \bibfield  {author} {\bibinfo {author} {\bibfnamefont {D.}~\bibnamefont {{Garfinkle}}}, \bibinfo {author} {\bibfnamefont {G.~T.}\ \bibnamefont {{Horowitz}}},\ and\ \bibinfo {author} {\bibfnamefont {A.}~\bibnamefont {{Strominger}}},\ }\bibfield  {title} {\bibinfo {title} {{Erratum: ``Charged black holes in string theory'' [Phys. Rev. D 43, 3140 (1991)]}},\ }\href {https://doi.org/10.1103/PhysRevD.45.3888} {\bibfield  {journal} {\bibinfo  {journal} {\prd}\ }\textbf {\bibinfo {volume} {45}},\ \bibinfo {pages} {3888} (\bibinfo {year} {1992})}\BibitemShut {NoStop}%
\bibitem [{\citenamefont {{Pugliese}}\ \emph {et~al.}(2011)\citenamefont {{Pugliese}}, \citenamefont {{Quevedo}},\ and\ \citenamefont {{Ruffini}}}]{2011PhRvD..83b4021P}%
  \BibitemOpen
  \bibfield  {author} {\bibinfo {author} {\bibfnamefont {D.}~\bibnamefont {{Pugliese}}}, \bibinfo {author} {\bibfnamefont {H.}~\bibnamefont {{Quevedo}}},\ and\ \bibinfo {author} {\bibfnamefont {R.}~\bibnamefont {{Ruffini}}},\ }\bibfield  {title} {\bibinfo {title} {{Circular motion of neutral test particles in Reissner-Nordstr{\"o}m spacetime}},\ }\href {https://doi.org/10.1103/PhysRevD.83.024021} {\bibfield  {journal} {\bibinfo  {journal} {\prd}\ }\textbf {\bibinfo {volume} {83}},\ \bibinfo {eid} {024021} (\bibinfo {year} {2011})},\ \Eprint {https://arxiv.org/abs/1012.5411} {arXiv:1012.5411 [astro-ph.HE]} \BibitemShut {NoStop}%
\bibitem [{\citenamefont {{Sen}}(1992)}]{1992PhRvL..69.1006S}%
  \BibitemOpen
  \bibfield  {author} {\bibinfo {author} {\bibfnamefont {A.}~\bibnamefont {{Sen}}},\ }\bibfield  {title} {\bibinfo {title} {{Rotating charged black hole solution in heterotic string theory.}},\ }\href {https://doi.org/10.1103/PhysRevLett.69.1006} {\bibfield  {journal} {\bibinfo  {journal} {\prl}\ }\textbf {\bibinfo {volume} {69}},\ \bibinfo {pages} {1006} (\bibinfo {year} {1992})},\ \Eprint {https://arxiv.org/abs/hep-th/9204046} {arXiv:hep-th/9204046 [hep-th]} \BibitemShut {NoStop}%
\bibitem [{\citenamefont {{Misner}}\ \emph {et~al.}(1973)\citenamefont {{Misner}}, \citenamefont {{Thorne}},\ and\ \citenamefont {{Wheeler}}}]{1973grav.book.....M}%
  \BibitemOpen
  \bibfield  {author} {\bibinfo {author} {\bibfnamefont {C.~W.}\ \bibnamefont {{Misner}}}, \bibinfo {author} {\bibfnamefont {K.~S.}\ \bibnamefont {{Thorne}}},\ and\ \bibinfo {author} {\bibfnamefont {J.~A.}\ \bibnamefont {{Wheeler}}},\ }\href@noop {} {\emph {\bibinfo {title} {{Gravitation}}}}\ (\bibinfo  {publisher} {{W. H. Freeman and Company}},\ \bibinfo {year} {1973})\BibitemShut {NoStop}%
\bibitem [{\citenamefont {{Bokhari}}\ \emph {et~al.}(2020)\citenamefont {{Bokhari}}, \citenamefont {{Rayimbaev}},\ and\ \citenamefont {{Ahmedov}}}]{2020PhRvD.102l4078B}%
  \BibitemOpen
  \bibfield  {author} {\bibinfo {author} {\bibfnamefont {A.~H.}\ \bibnamefont {{Bokhari}}}, \bibinfo {author} {\bibfnamefont {J.}~\bibnamefont {{Rayimbaev}}},\ and\ \bibinfo {author} {\bibfnamefont {B.}~\bibnamefont {{Ahmedov}}},\ }\bibfield  {title} {\bibinfo {title} {{Test particles dynamics around deformed Reissner-Nordstr{\"o}m black hole}},\ }\href {https://doi.org/10.1103/PhysRevD.102.124078} {\bibfield  {journal} {\bibinfo  {journal} {\prd}\ }\textbf {\bibinfo {volume} {102}},\ \bibinfo {eid} {124078} (\bibinfo {year} {2020})}\BibitemShut {NoStop}%
\end{thebibliography}
\end{document}